\documentclass[a4paper,10pt]{article}
\usepackage{amsmath,amssymb,mathtools}  
\usepackage{soul}                       
\usepackage[usenames,dvipsnames]{xcolor}
\usepackage[raggedright]{titlesec}
\usepackage{scrextend}
\deffootnote{2em}{0em}{\thefootnotemark\quad}
\usepackage{multicol}
\usepackage[left=0.75in,right=0.75in,top=1.0in,bottom=1.25in]{geometry}
\usepackage{hyperref}
\usepackage{float}
\usepackage{xcolor}
\usepackage{booktabs}
\usepackage{caption}
\expandafter\let\csname equation*\endcsname\relax
\expandafter\let\csname endequation*\endcsname\relax

\hypersetup{
  colorlinks   = true, 
  urlcolor     = blue, 
  linkcolor    = blue, 
  citecolor   = blue 
}

\definecolor{seablue}{rgb}{0.0, 0.4784, 0.6471}
\definecolor{deepskyblue}{rgb}{0.0, 0.75, 1.0}
\newcommand{\dd}{\text{d}}

\newcommand{\dx}{\text{d}x}

\setstcolor{red}    
\setulcolor{red}    
\allowdisplaybreaks 

\let\oldabstract\abstract
\let\oldendabstract\endabstract
\makeatletter
\renewenvironment{abstract}
{%
               {\list{}{\addtolength{\leftmargin}{4em} 
                        \listparindent 1.5em%
                        \itemindent    \listparindent%
                        \rightmargin   \leftmargin%
                        \parsep        \z@ \@plus\p@}%
                \item\relax}%
               {\endlist}%
\oldabstract}
{\oldendabstract}
\makeatother

\title{\LARGE\textbf{\textsf{On the KG-constrained general disformal transformation\\ of the Einstein-Hilbert action}}}

\author{\normalsize Allan L. Alinea\footnote{main and corresponding author: alalinea@up.edu.ph} \;and Joshwa DJ. Ordo\~nez}

\date{}

\begin{document}

\maketitle
\vspace{-2.25em}
\noindent
\begin{center}
{\small Astrophysics, Particle Physics, and Nuclear Physics Research Cluster\\Physics Division, Institute of Mathematical Sciences and Physics\\ University of the Philippines Los Ba\~nos\\4031 College, Los Ba\~nos, Laguna, Philippines}
\end{center}

\bigskip
\begin{abstract}
\noindent
Although with great successes in explaining phenomena and natural behaviour involving the Universe or a part thereof, the General Theory of Relativity is far from a complete theory. Focusing on its extension within the framework of scalar tensor theory, we investigate the transformation of the Einstein-Hilbert action under the general disformal transformation coupled with a set of triple KG constraints. The idea is that the general disformal transformation leads to an extremely complicated action necessitating a set of constraints to tame the resulting form. Motivated by previous studies on the invertibility and invariance of the massless Klein-Gordon equation under the general disformal transformation, we identify three constraints that significantly simplifies the transformed Einstein-Hilbert action. In four spacetime dimensions, we find that the transformed action is a sum of the original action and a disformal contribution involving a six-term Lagrangian that includes the Ricci tensor coupled to a sum of derivatives of scalar fields and kinetic terms. In three spacetime dimensions, the disformal contribution becomes a five-term Lagrangian. Lastly, in two spacetime dimensions, the Einstein-Hilbert action is invariant under the constrained disformal transformation.

\par
\vspace{0.75em}
\noindent
{\textbf{keywords:} \textit{disformal transformation, Einstein-Hilbert action, Klein-Gordon equation, \\ \phantom{dominatedxx}Horndeski action, scalar-tensor theory}}	
\end{abstract}

\bigskip
\begin{multicols}{2}
\section{Introduction}
\label{secIntro}
\noindent
\textit{Disformal transformation} is a metric transformation generalising the more familiar \textit{conformal transformation} in general relativity. Whereas the conformal transformation involves scaling of the metric, $ g_{\mu \nu } $, by a (scalar) field-dependent conformal factor, $A(\phi )$, that is, $g_{\mu \nu } \rightarrow \widehat g_{\mu \nu } =  A(\phi ) g_{\mu \nu } $, the original disformal transformation due to Bekenstein \cite{Bekenstein:1992pj}, includes a direction-dependent variation of the metric in the transformation. In particular, $ \widehat g_{\mu \nu } = A (\phi,X) g_{\mu \nu } + B(\phi ,X)\phi _{;\mu }\phi _{;\nu }$, where $ B $ is the disformal factor and the semicolon `;' denotes covariant derivative; \textit{i.e.}, $\phi_{;\mu} = \nabla_\mu\phi$. The disformal part, $  B(\phi ,X)\phi _{;\mu }\phi _{;\nu }$, serves to deform the metric along the direction of variation of the scalar field. In addition to this, the generalisation includes dependence on the kinetic term, $X \equiv -(1/2)g^{;\mu \nu }\phi _{;\mu }\phi _{;\nu }$, which also happens to encode the (inverse) metric. In perspective then, the Bekenstein's disformal transformation is a two-fold generalisation of the usual conformal transformation.

Because of this, it is not straightforward to have a relatively simple physical intuition about the disformal transformation beyond our concise description above, in contrast to that for the conformal transformation. Nonetheless, we can take a few steps forward in its physical interpretation\footnote{Readers interested in a more thorough discussion of the physical interpretation of, constraints on, and value of the disformal transformation in relation to the conformal transformation and the more general Scalar-tensor theories, may refer to Ref. \cite{Bettoni:2013diz} and the references therein. For numerous applications of disformal transformation in relation to Horndeski action, see Ref. \cite{Kobayashi:2019hrl}.} by switching off the dependence on $ X $ above to yield the \textit{special} disformal transformation; \textit{i.e.,} $ \widehat g_{\mu \nu } = A (\phi) g_{\mu \nu } + B(\phi)\phi _{;\mu }\phi _{;\nu }$. When $ B =0 $, the conformal transformation scales the metric but preserves the angles between tangent vectors on a manifold that the metric describes. Physically, this means that the light cones associated with the manifold remain the same modulo scaling and the underlying causal structure is preserved. When $ B \ne 0 $, the disformal transformation can widen or narrow the light cones. Certainly, the latter possibility can lead to undesirable acausal structures. However, with  suitable constraint(s) in the conformal and disformal factors (see Ref. \cite{Bettoni:2013diz}), disformal transformation can give rise to a richer set of structures where the light cones may `fold' within the bounds of causality.

Beyond this, when the conformal and disformal factors are no longer dependent on $ \phi $ alone, and additional kinetic-term dependencies may come into play, we are welcomed with far richer set of possible structures and flexibility that are not found in conformal transformation alone. Bekenstein, who introduced disformal transformation, exploited this greater `flexibility' in the 1990s to relate the physical metric describing Riemannian geometry, to the gravitational metric, within the context of Finslerian geometry \cite{Bekenstein:1992pj}. Since then, unbounded by the limitations of conformal transformation alone, disformal transformation found numerous applications beyond the initial framework within which the transformation was introduced. Within two decades after the publication of Bekenstein's work, there were several studies on symmetries of primordial cosmological perturbations\footnote{For studies on the conformal transformation involving primordial cosmological perturbation, interested readers may refer to Refs. \cite{Makino:1991sg,Fakir:1990eg,Alinea:2015pza,Alinea:2016qlf,Alinea:2017ncx}.} \cite{Tsujikawa:2014uza,Domenech:2015hka,Motohashi:2015pra,Alinea:2020laa} under disformal transformation, disformal and Higgs cosmic inflations \cite{Sato:2017qau,Kaloper:2003yf,Gialamas:2024jeb,Gialamas:2020vto,Gialamas:2021enw}, invariance of field equations of motion \cite{Falciano:2011rf,Alinea:2022ygr,Goulart:2013laa,Bittencourt:2015ypa}, variations of actions in scalar tensor theories \cite{Bettoni:2013diz,Kobayashi:2019hrl,Alinea:2020sei,Gleyzes:2014qga,Langlois:2015cwa}, etc. More recently, researches on disformal transformation involve theoretical investigations of wormholes \cite{Chatzifotis:2021hpg}, hairy black holes \cite{BenAchour:2020wiw}, mimetic gravity \cite{Domenech:2023ryc}, nonminimally coupled Proca theory \cite{Hell:2024xbv}, and ``most'' general action forms subject to invertibility in relation to Ostrogradsky instability \cite{Takahashi:2021ttd,Takahashi:2022mew,Woodard:2015zca}, to name a few.

Although we seem to have a plethora of distinct studies on disformal transformation, the nature of them may be divided into three interrelated categories, namely, (a) symmetry, (b) extension of an existing theory or action, and (c) consistency of the transformation in relation to the corresponding action. On the side of symmetry, we have primordial cosmological perturbations---quantum fluctuations that served as seeds of what we nowadays observe as galaxies and clusters thereof---that remain invariant under disformal transformation \cite{Tsujikawa:2014uza,Domenech:2015hka,Motohashi:2015pra,Alinea:2020laa}. Also included in this category, are the invariance of the equations of motion (under some conditions) such as Klein-Gordon equation \cite{Falciano:2011rf,Alinea:2022ygr}, Maxwell's equation \cite{Goulart:2013laa,Goulart:2020wkq}, and Dirac equation \cite{Domenech:2015hka}. For the extension of existing theories, disformal transformation is used to generate new actions, usually within the context of scalar-tensor theories, to possibly find new ways to address pressing problems such as dark matter and dark energy, or investigate new spacetime structures \cite{Bettoni:2013diz,Alinea:2020sei,Kobayashi:2019hrl,Gleyzes:2014qga,Langlois:2015cwa,Gleyzes:2014dya}. Lastly, along the way of this transformation, instabilities \cite{Woodard:2015zca} may be introduced in an otherwise healthy action describing a physical system. The third category involves limiting general disformal transformation or the action upon which it acts, in order to evade the existence of ghosts or prevent unbounded energy \cite{Takahashi:2021ttd,Takahashi:2022mew,BenAchour:2016cay,Galtsov:2018xuc}.

This study mainly falls under the second category involving the transformation of an action. In particular, we consider the simplest of the actions in the set of scalar tensor theories, at the heart of modern cosmology; namely, the Einstein-Hilbert action\footnote{In this work, we use the reduced Planck units, where $8\pi G = \hbar = c = 1$, and the metric signature, $(-+++)$.}:
\begin{align}
    \label{ehaction}
    S = \frac{1}{2}\int \dx^n\,\sqrt{-g}\, R.
\end{align}
Here, $R$ is the Ricci scalar, $g$ is the determinant of the metric, $g_{\mu\nu}$, and $n$ is the number of spacetime dimensions. We aim to find its variation under the disformal transformation \cite{Alinea:2020laa,Alinea:2022ygr,Takahashi:2021ttd} given by\footnote{For a closely related study involving Bekenstein's disformal transformation, see Ref. \cite{Alinea:2024jrf}.}
\begin{align}
\label{genDT}
g_{\mu\nu}
\rightarrow 
\widehat g_{\mu\nu}
=
Ag_{\mu\nu} + \Phi_\mu\Phi_\nu,
\end{align}
where $\Phi_\mu \equiv C\phi_{;\mu} + DX_{;\nu}$, with the semicolon `;' denoting covariant derivative; that is, $\phi_{;\mu} = \nabla_\mu\phi$. The functionals $A$ and $(C,D)$ are called the conformal factor and (a pair of) disformal factors, respectively. In general, they all depend on the scalar field $\phi$ and $(X, Y, Z)$, where 
\begin{align*}
X \equiv -\frac{1}{2}g^{\mu\nu} \phi_{;\nu}\phi_{;\nu},\;
Y \equiv g^{\mu\nu}\phi_{;\mu}X_{;\nu},\;
Z \equiv g^{\mu\nu}X_{;\mu}X_{;\nu}.
\end{align*}
Such a form of transformation is a generalisation of the original \textit{Bekenstein's disformal transformation}; that is, $
\widehat g_{\mu\nu} = A(\phi, X) g_{\mu\nu} + B(\phi, X) \phi_{;\mu} \phi_{;\nu}$ \cite{Bekenstein:1992pj}. The equation given by (\ref{genDT}) is an umbrella transformation covering those specific cases investigated in Refs. \cite{Tsujikawa:2014uza,Domenech:2015hka,Motohashi:2015pra,Bettoni:2013diz,Minamitsuji:2014waa} in the study of primordial cosmological perturbations and extensions of the Horndeski action \cite{Horndeski:1974wa} in scalar-tensor theory.

It is worth noting that this study is the second in line involving the disformal transformation of the Einstein-Hilbert action. Inspired by the form invariance of the Horndeski action\footnote{The Horndeski action is the most general scalar-tensor action that yields second order equations of motion without the need for auxiliary conditions in addition to the action itself.} under the \textit{special disformal transformation}, $\widehat g_{\mu\nu} = A(\phi)g_{\mu\nu} + B(\phi)\phi_{;\mu}\phi_{;\nu}$, studied in Ref. \cite{Bettoni:2013diz}, we investigated in Ref. \cite{Alinea:2020sei} the curious case of the Einstein-Hilbert action upon the same transformation. Rather informally, the idea in Ref. \cite{Bettoni:2013diz} is that in spite of the complicated four-term sub-Lagrangians constituting the Horndeski action, these sub-Lagrangians ``conspire'' on special disformal transformation to preserve the original form of the action; meaning, the result is simply another Horndeski action. Given that the Einstein-Hilbert action is a special case of the Horndeski action, it is interesting to know the consequence of the special disformal transformation on this part alone of the Horndeski action; that is, without all other sub-Lagrangians to ``conspire'' with, to possibly prevent unstable terms. As it turns out, the result in Ref. \cite{Alinea:2020sei} is an action still perfectly fitting the Horndeski form sans the fifth sub-Lagrangian involving the Einstein tensor and cubic d'Alembertians of the scalar field.

For the current work, we consider a much more general disformal transformation given by (\ref{genDT}). Given this level of generality, however, one would expect a lengthy and complicated transformed Einstein-Hilbert action. Indeed, following the trajectory of derivation laid down in Ref. \cite{Alinea:2020sei}, the covariant derivative terms involving the transformed Christoffel symbols coupled with the inverse disformal metric (see Sec. \ref{secdtrictens}) in the equation for the transformed Ricci tensor, leads to an extremely lengthy expression with numerous high-order derivative terms. Such `state-of-affairs' call for some constraints that could be imposed on the action itself or the functionals involved in the disformal transformation, as is usually done in scalar-tensor theory, to at least simplify the resulting action and possibly remove some undesirable terms. To this end, for the one-term Lagrangian of the Einstein-Hilbert action, we choose to impose conditions on the conformal and disformal functionals $A, C,$ and $D$ in the general disformal transformation given by (\ref{genDT}).

Having said this, we see more than one way to impose conditions on the functionals in the general disformal transformation above if the aim is a relatively `simple' resulting action; that is, one with small number of terms and possible instabilities. Knowing the time and space limitations in this paper, we find it prudent to use one good set of constraints and leave other possible sets of constraints for future studies. In this study, we look no further and employ the  constraints in our earlier work in Ref. \cite{Alinea:2022ygr} involving the massless Klein-Gordon equation\footnote{This work was inspired by Ref. \cite{Falciano:2011rf} where the massless Klein-Gordon equation was found to be invariant under the (less general) \textit{Bekenstein's disformal transformation} under some constraint on the disformal factor.}. Based on this study, the massless Klein-Gordon equation for scalar fields and correspondingly, the Klein-Gordon action, remain invariant under the general disformal transformation given by (\ref{genDT}) if the disformal factors $C$ and $D$ take on a special form in relation to the conformal factor $A$; see (\ref{orthocond}), (\ref{cdconstr}), and (\ref{invconstr}) in Sec. \ref{seckgdisf} below.

The idea for this choice of set of constraints in the current work is at least two-fold. First, the Klein-Gordon action and the Einstein-Hilbert action both belong as (two distinct) special cases of the Horndeski action \cite{Horndeski:1974wa}. As such, it is interesting to know if the general disformal transformation of the Einstein-Hilbert action, subject to the same constraints, is invariant or at least reduces to a subset of the Horndeski action, similar to that in Ref. \cite{Alinea:2020sei}. In hindsight, even if the second scenario is not completely attainable due to the presence of a few higher order derivative terms, we would find it  odd to have two `subsets' of the Horndeski action to have different constrained (symmetry) transformations\footnote{We leave it for future work to investigate such a possible unusual scenario.}. Second, the constrained disformal transformation in Ref. \cite{Alinea:2022ygr} that yields invariant massless Klein-Gordon equation is invertible, in general\footnote{Interested readers may refer to Ref. \cite{Takahashi:2022mew} for a more general treatment of invertibility.}. Consequently, using the same constrained transformation frees us from the problem of invertibility (of the transformation itself) that could lead to additional possible instabilities and issues involving ghost terms and the primordial cosmological perturbations, within the much larger framework of Horndeski theory and extensions thereof.

With these justifications, we are set to derive the variation of the Einstein-Hilbert action under the general disformal transformation given by (\ref{genDT}) subject to the same constraints for the invariant massless Klein-Gordon equation in Ref. \cite{Alinea:2022ygr}. In line with this, the paper is organised as follows. In Sec. \ref{seckgdisf}, we present a short review of the disformal transformation of the massless Klein-Gordon equation based on Ref. \cite{Alinea:2022ygr} and some ideas from its `parent' paper namely, Ref. \cite{Falciano:2011rf}. The objective is to put forward the explicit forms of the constraint on the disformal factors $(C,D)$ in relation to the conformal factor $A$, from a solid and concise ground. After this, we delve into the derivation of the disformal Ricci tensor in Sec. \ref{secdtrictens}, starting from the disformal metric and its inverse. This section is mostly mathematical `acrobatics' but can provide good insights about the complexity of the resulting transformed Einstein-Hilbert action in the absence of constraints on the disformal factors. In Secs. \ref{secdtehact} and \ref{secDimlim}, we present the main results of this work and some insights on the massless KG-constrained disformal transformation of the Einstein-Hilbert action. Lastly, we give our concluding remarks and future prospects in Sec. \ref{seConclude}.

\section{Invariance of the massless Klein-Gordon equation under disformal transformation}
\label{seckgdisf}
\noindent
The massless Klein-Gordon equation is, in general, \textit{not} invariant under the conformal transformation in $n$ spacetime dimensions; the exception\footnote{To wit, the $\widehat \square \phi = \frac{\square \phi}{A} + (n-2)\frac{\phi^{;\alpha}A_{;\alpha}}{2A^2},$ under $g_{\mu\nu}\rightarrow\widehat g_{\mu\nu} = A(\phi)g_{\mu\nu}$.} is for $n=2$ dimensions. Interestingly, it is invariant under the Bekenstein's disformal transformation, $\widehat g_{\mu\nu} = A(\phi,X)g_{\mu\nu} + B(\phi,X)\phi_{;\mu}\phi_{;\nu}$, subject to a constraint relating the disformal factor, $B$, to the conformal factor, $A$. Symbolically, 
\begin{align}
\widehat \square \phi = 0
\quad\Rightarrow\quad
\square \phi = 0,
\end{align}
where $\square \equiv g^{\mu\nu}\nabla_\mu \nabla_\nu $ and $\widehat \square \equiv \widehat g^{\mu\nu}\widehat\nabla_\mu \widehat\nabla_\nu $, with $\widehat g^{\mu\nu}$ being the inverse disformal metric. This is under the provision\footnote{We are using here the notation in Ref. \cite{Alinea:2022ygr} and its slightly more general result compared to Ref \cite{Falciano:2011rf}, where $b^2 = 1$.} that \cite{Falciano:2011rf,Alinea:2022ygr},
\begin{align}
B(\phi,X)
=
\frac{A - b^2 A^{n-1}}{2X}
\quad
(b = \text{const.})
\end{align}

When the Bekenstein's disformal transformation is generalised to that given by (\ref{genDT}) above \cite{Alinea:2020laa,Alinea:2022ygr,Takahashi:2021ttd}, the transformation of $\square \phi$ is complicated by the four functional dependencies $(\phi, X, Y, Z)$ and an additional disformal factor. In this section, we present a concise derivation of the transformation of $\square \phi$ under the transformation (\ref{genDT}) based mainly on the logic in Ref. \cite{Alinea:2022ygr}. Along the way of calculation we identify the constraints relating the disformal factors to each other and to the conformal factor in the metric $\widehat g_{\mu\nu}$. Moreover, we limit the functional dependency of the conformal factor for the disformal transformation to be invertible; that is, $g_{\mu\nu}$ being expressible in terms of the disformally transformed (hatted) quantities.

To proceed, one may start with the expansion of the hatted double covariant derivatives in the expression for $\widehat \square \phi$ as $\widehat \nabla _\mu \widehat \nabla _\nu \phi =	\phi _{;\nu ,\mu } -	\widehat \Gamma ^\alpha_{\mu \nu }\phi _{;\alpha },$ where the hatted Christoffel symbol, $\widehat \Gamma^\alpha_{\mu\nu},$ simply follows the form of the original one, $\Gamma^\alpha_{\mu\nu}$; that is, 
\begin{align}
    \label{whgamma}
    \Gamma^\alpha_{\mu\nu}
    &=
    \tfrac{1}{2} g^{\alpha\beta}(
        g_{\beta\mu,\nu}
        +
        g_{\nu\beta,\mu}
        -
        g_{\mu\nu,\beta}
    ),
    \nonumber
    \\[0.5em]
    \widehat \Gamma^\alpha_{\mu\nu}
    &=
    \tfrac{1}{2}\widehat g^{\alpha\beta}(
        \widehat g_{\beta\mu,\nu}
        +
        \widehat g_{\nu\beta,\mu}
        -
        \widehat g_{\mu\nu,\beta}
    ).     
\end{align}
Using the Sherman-Morrison formula,  the inverse disformal metric can be written as\footnote{It is straightforward to show that $\widehat g^{\mu\alpha}\widehat g_{\alpha \nu} = \delta^\mu_\nu$.}
\begin{align}
    \label{invg}
	\widehat g^{\mu \nu }
	&=
	\frac{g^{\mu \nu }}{A}
	-
	\frac{\Phi ^{\mu} \Phi ^{\nu} }{A(A - 2\chi)} ;
	\quad
	(\chi \equiv -\tfrac{1}{2}g^{\mu \nu }\Phi _\mu \Phi _\nu  ).
\end{align}

Then, given the disformal metric and its inverse, the hatted Christoffel symbol can be decomposed as $\widehat \Gamma^\alpha _{\mu \nu } = \Gamma ^\alpha _{\mu \nu } + F^\alpha _{\mu \nu }$, where the last term is a \textit{tensorial} contribution from the general disformal transformation, involving derivatives of the conformal and disformal factors; specifically,
\begin{align}
    \label{Fmunu}
    F^\alpha _{\mu \nu }
	&=
	\frac{
		A_{;\nu }\delta ^\alpha _\mu 
		+
		A_{;\mu }\delta ^\alpha _\nu 
		-
		A^{;\alpha }g_{\mu \nu }	
	}{2A }
 	+
	\frac{\Phi^\alpha\Phi _{(\mu;\nu )}}{A - 2\chi } 
    \\[0.5em]
	&\qquad
    -\,
	\frac{
		\Phi ^\alpha(
			A _{;\nu }\Phi _\mu 
			+
			A _{;\mu }\Phi _\nu 
			-
			A _{;\beta }\Phi ^\beta g_{\mu \nu }
		)
	}{2A (A - 2\chi )} 
	\nonumber
	\\[0.5em]    
	&\qquad
	+\,
	\left[
		\frac{g^{\alpha \beta }}{A}
		-
		\frac{\Phi ^\alpha \Phi ^\beta }{A(A - 2\chi )} 
	\right]
	(
		\Phi _\mu\Phi _{[\beta;\nu ]} 
		+
		\Phi _{\nu }\Phi _{[\beta;\mu]} 	
	),
    \nonumber
\end{align}
where 
\begin{align}
    \Phi_{(\mu;\nu)} 
    &\equiv 
    \tfrac{1}{2}(\Phi_{\mu;\nu} + \Phi_{\nu;\mu})
    \quad\nonumber\\
    \text{and} \quad
    \Phi_{[\mu;\nu]} 
    &\equiv 
    \tfrac{1}{2}(\Phi_{\mu;\nu} - \Phi_{\nu;\mu}).
\end{align}
Such a decomposition of the Christoffel symbol leads us to $\widehat \nabla _\mu \widehat \nabla _\nu \phi = \phi _{;\mu \nu } - F^\alpha  _{\mu \nu }\phi _{;\alpha }$. Since $F^\alpha _{\mu \nu }$ is symmetric with respect to the indices $(\mu,\nu)$ then $\widehat \nabla _\mu \widehat \nabla _\nu \phi$ is symmetric with respect to the same pair of indices, as the original $\phi_{;\mu\nu}$ (in torsionless manifold).

To find $\widehat \square \phi$ we  need to contract $\widehat \nabla _\mu \widehat \nabla _\nu \phi$ with the inverse disformal metric. Upon using the equation above for the hatted Christoffel symbol and (\ref{invg}) we find
\begin{align}
    \label{hatdAlem}
	\widehat \square \phi 
 =
 \frac{\square\phi }{A}
	-
	\frac{g^{\mu \nu} F^\alpha _{\mu \nu }\phi _{;\alpha }}{A}  
	-
	\frac{\Phi ^{\mu} \Phi ^{\nu} \phi _{;\mu \nu }}{A(A - 2\chi)} 
	+
	\frac{\Phi ^{\mu} \Phi ^{\nu} F^\alpha _{\mu \nu }\phi _{;\alpha }}{A(A - 2\chi)}.
\end{align}
Clearly, for the massless Klein-Gordon equation to be invariant under the general disformal transformation (\ref{genDT}), $\widehat \square \phi $ must be proportional to $\square \phi$. Furthermore, all those terms not proportional to $\square \phi$ in the last three terms on the right hand side should sum up to zero. Knowing the complicated form of $F^\alpha_{\mu\nu}$ given by (\ref{Fmunu}), this can be a challenging task, unless we have the trivial scenario where both the disformal factors vanish and the conformal factor is unity.

Having said this, we observe upon expansion of the last three terms on the right hand side of (\ref{hatdAlem}) that only the last one generates $\square X$. More specifically, 
\begin{align}
	\phi _{;\alpha }g^{\mu \nu }F^\alpha _{\mu \nu }	
	&=
	\frac{
		(2-n)\phi ^{;\alpha }A_{;\alpha }
	}{2A }
	-
	\frac{
		(2-n)\phi _{;\alpha }\Phi ^\alpha\,
		A _{;\beta  }\Phi ^\beta 
	}{2A (A - 2\chi )}
    \nonumber
    \\[0.5em]
    &\qquad
	+\,
	\frac{2\phi ^{;\alpha  }\Phi ^\beta \Phi _{[\alpha ;\beta  ]}}{A}
	+
	\frac{\phi _{;\alpha }\Phi^\alpha \Phi^\beta {}_{;\beta  }}{A - 2\chi },
\end{align}
with ${\Phi ^\beta }_{;\beta } = C_{;\beta }\phi ^{;\beta } + C\square \phi + D_{;\beta }X^{;\beta } + D\square X$. If $\widehat \square \phi$ is then to be proportional to $\square \phi$, one may at least zero out the term involving $\square X$ by imposing the `orthogonality' condition $\phi^{;\alpha} \Phi_{\alpha}$ = 0. To be clear, such a condition is not an equation for the scalar field $\phi$ in addition to the Klein-Gordon equation. Rather it should be seen as a constraint binding the disformal factors $C$ and $D$; in particular, $2CX = DY$.

Fortunately, the application of the `orthogonality' condition in (\ref{hatdAlem}) not only removes terms involving $\square X$ but also all terms involving $\square \phi$ apart from the first term on the right hand side. However, this constraint is not enough to make the massless Klein-Gordon equation  invariant under the disformal transformation (\ref{genDT}). Specifically, when this constraint is applied, the equation for $\widehat \square \phi$ given by (\ref{hatdAlem}) becomes
\begin{align}
    \label{dAlemOrtho}
	\widehat \square \phi 
	&=
	\frac{\square\phi }{A}
	+
	\frac{[n(A - 2\chi) -2A + 6\chi ]\phi ^{;\alpha }A_{;\alpha }}
	{2A^2(A - 2\chi)}
    \nonumber
    \\[0.5em]
	&\quad
    -\,
	\frac{\Phi ^{\mu} \Phi ^{\nu} \phi _{;\mu \nu }}{A(A - 2\chi)}
	-
	\frac{2\phi ^{;\alpha  }
		\Phi ^\beta \Phi _{[\alpha ;\beta  ]}}
	{A(A - 2\chi )}.
\end{align}
To attain the desired relation, namely, $\widehat \square \phi \sim \square \phi$, all the last three terms on the right hand side should then sum up to zero.

To this end, one can decompose the disformal factor, $C^2 = B$, as a product of three functionals, $(q, h, f)$, with $f$ being a functional of the conformal factor, $A$.
\begin{align}
    B = q(X,Y,Z) h(X) f(A).
\end{align}
Using this form of $B$ in the expansion of the last three terms on the right hand side of (\ref{dAlemOrtho}) and correspondingly grouping terms of similar forms to make each group vanish (e.g., coefficients of $\phi^{;\alpha}Y_{;\alpha}$ and $\phi^{;\alpha}Z_{;\alpha}$ are set to zero) leads to 
\begin{align}
    q &= \frac{Y^2}{Y^2 + 2XZ},
    \nonumber
    \\[0.5em]
    0 &= h + Xh_X,
    \nonumber
    \\[0.5em]
    0 &= 2A f_A	+ 2f(n - 3)+ (n-2)A,
\end{align}
where $f_A = \dd f/\dd A$ and $h_X = \dd h /\dd X$. Upon solving the last two differential equations above, we find 
\begin{align}
	\label{Bxdisft}
	B
	=
	-\frac{Y^2(A - b^2A^{3 - n})}{2X(Y^2 + 2XZ)},
\end{align}
where $b^2 = \text{constant}$. This results in $\widehat \square \phi = (\square \phi)/A$.

We therefore conclude that the massless Klein-Gordon equation in $n$ spacetime dimensions is invariant under the disformal transformation (\ref{genDT}), subject to two constraints, namely, (a) the `orthogonality' condition, $\phi^{;\alpha}\Phi_\alpha = 0,$ binding the disformal factors as $2CX = DY$, and (b) the equation relating $B = C^2$ to the conformal factor as given by (\ref{Bxdisft}). It is worth noting that both disformal factors have infinitely many possible forms\footnote{In two dimensions, $n = 2$, with $b^2 = 1$ the two disformal factors become trivial and we have a purely conformal transformation.}, albeit limited by their relationship to each other and to the conformal factor. On the other hand, the conformal factor, $A = A(\phi, X, Y, Z)$, seems to be truly `free' for the disformal invariance of the massless Klein-Gordon equation under (\ref{genDT}) to hold. 

Having said this, going beyond the disformal invariance of the massless Klein-Gordon equation, we find that if the disformal transformation (\ref{genDT}) under the two mentioned constraints is to be invertible, then some dependencies of $A$ must be removed. We learned in Ref. \cite{Alinea:2022ygr} that the invertibility condition on $\widehat g_{\mu\nu}(g_{\mu\nu},\phi,X,Y,Z)$ leads to a `ladderised' sequence of functional dependencies as 
\begin{align}
    \widehat X 
    &= 
    \widehat X(\phi, X),
    \nonumber
    \\[0.5em]
    \widehat Y 
    &= 
    \widehat Y(\phi, X, Y),
    \nonumber
    \\[0.5em]
    \widehat Z 
    &= 
    \widehat Z(\phi, X, Y, Z),  
\end{align}
with the requirement that the conformal factor be limited\footnote{Although $A = A(\phi,X)$, the disformal factors retain dependency on $(\phi, X, Y, Z)$.} to $A = A(\phi,X)$. This allows us to solve, at least in principle, for $X = X(\phi, \widehat X),\, Y = Y(\phi, \widehat X, \widehat Y),\, Z = Z(\phi, \widehat X, \widehat Y, \widehat Z)$, and finally, $ g_{\mu\nu}(\widehat g_{\mu\nu},\widehat \phi,\widehat X,\widehat Y,\widehat Z)$. Surprisingly, with this auxiliary condition on the conformal factor, the `orthogonality' condition, $2CX = DY$, in the original frame takes the same form in the hatted frame; that is, $2\widehat C\widehat X = \widehat D\widehat Y$.

To end this section, we list down the three conditions we impose for the invariance of the massless Klein-Gordon equation. Each condition or constraint is also given name for easy reference later. Taking their root from the Klein-Gordon equation, we call the triple constraints as the KG constraints or conditions from hereon:
\begin{itemize}
    \item 
    `orthogonality' condition:
    \begin{align}\label{orthocond}\Phi^\mu \phi_{;\mu} = 0,\end{align}
    \item
    conformal-disformal constraint:
    \begin{align}\label{cdconstr}B
	=
	\displaystyle-\frac{Y^2(A - b^2A^{3 - n})}{2X(Y^2 + 2XZ)},\end{align}
    \item
    invertibility constraint:
    \begin{align}\label{invconstr}A = A(\phi,X).\end{align}
\end{itemize}
We use these set of KG constraints for the disformal transformation (\ref{genDT}) of the Einstein-Hilbert action.

\section{KG-Constrained disformal transformation of the Ricci tensor}
\label{secdtrictens}

\noindent 
In this section, we derive the transformed Ricci tensor under the general disformal transformation given by (\ref{genDT}) and constrained by the KG conditions specified in the immediately preceding section. To lessen extremely long sub-expressions in the derivation, the conditions are applied as soon as we find an opportunity to apply them. Our pathway of derivation follows Ref. \cite{Alinea:2020sei} on the \textit{special} disformal transformation of the Einstein-Hilbert action; that is, $\widehat g_{\mu\nu} = A(\phi) g_{\mu\nu} + B(\phi) \phi_{;\mu}\phi_{;\nu}$. The idea, in short, is to start with the expression for the hatted Riemann curvature tensor in terms of the hatted Christoffel symbol, and then contract respective indices to gain the hatted Ricci tensor. Since the hatted Christoffel symbol, decomposes as a sum of the original (unhatted) Christoffel symbol and terms contributed by the general disformal transformation (see Sec. \ref{seckgdisf}), we expect an analogous decomposition for the hatted Ricci tensor.

The disformally transformed Riemann curvature tensor can be written following its original (unhatted) form as
\begin{align}
	\widehat R^\alpha {}_{\mu \beta \nu }
	&=
	-\widehat \Gamma ^\alpha _{\mu \beta ,\nu }
	+
	\widehat \Gamma ^\alpha _{\mu \nu ,\beta }
	-
	\widehat \Gamma ^\rho _{\mu \beta }\widehat \Gamma ^\alpha _{\rho \nu }
	+
	\widehat \Gamma ^\rho _{\mu \nu }\widehat \Gamma ^\alpha _{\rho \beta },
\end{align}
where the hatted Christoffel symbols are expressed in terms of the hatted disformal metric and its inverse; see the second of (\ref{whgamma}). To find the hatted Ricci tensor, we simply contract the indices, $\alpha$ and $\beta$.
\begin{align}
	\label{hattedRicci}
	\widehat R_{\mu \nu }
	=
	-\widehat \Gamma ^\alpha _{\mu \alpha ,\nu }
	+
	\widehat \Gamma ^\alpha _{\mu \nu ,\alpha }
	-
	\widehat \Gamma ^\rho _{\mu \alpha }\widehat \Gamma ^\alpha _{\rho \nu }
	+
	\widehat \Gamma ^\rho _{\mu \nu }\widehat \Gamma ^\alpha _{\rho \alpha }.
\end{align}
Because the Christoffel symbol decomposes as $\widehat \Gamma^\alpha _{\mu \nu } = \Gamma ^\alpha _{\mu \nu } + F ^\alpha _{\mu \nu }$, where $F ^\alpha _{\mu \nu }$ is given by (\ref{Fmunu}), the hatted Ricci tensor takes the form given by 
\begin{align}
	\widehat R_{\mu \nu }
	=
	R_{\mu \nu }
	-
	F^\alpha _{\mu \alpha ;\nu }
	+
	F^\alpha _{\mu \nu ;\alpha }
	-
	F^\rho _{\mu \alpha }F^\alpha _{\rho \nu }
	+
	F^\rho _{\mu \nu }F^\alpha _{\rho \alpha }.
    \nonumber
\end{align}
In other words, the disformally transformed Ricci tensor is a sum of the original Ricci tensor and  terms due to the general disformal transformation. Observe that even with these disformal contributions, the Ricci tensor retains its tensorial nature and symmetry with respect to its indices.

The next step in our derivation is the calculation of the disformal contributions in the equation above for $\widehat R_{\mu\nu}$, namely,
\begin{align} 
    \label{dcontribrmn}
    \mathcal D_{\mu\nu}
    \equiv
	-
	F^\alpha _{\mu \alpha ;\nu }
	+
	F^\alpha _{\mu \nu ;\alpha }
	-
	F^\rho _{\mu \alpha }F^\alpha _{\rho \nu }
	+
	F^\rho _{\mu \nu }F^\alpha _{\rho \alpha }
\end{align}
Of these contributions, the first two terms involve covariant derivatives of $F^\alpha_{\mu\nu}$. For  this reason, they may involve undesirable third order derivative terms in the field variables $(\phi, X, Y, Z)$. In this section, we write down their explicit forms to identify these terms with the insight to `massage' them in the following section to produce terms involving at most second order derivatives. On the other hand, the last two terms are essentially a lengthy product expansion of the contraction of two $F^\alpha _{\mu \nu}$'s. It is clear from (\ref{Fmunu}) that these contractions go as far as second order derivative only in the field variables $(\phi, X, Y, Z)$. As such, we simply `encode' them in the final equation for $R_{\mu\nu}$ without explicitly writing them down separately in this section.

Starting with $F^\alpha_{\mu\alpha;
\nu}$ in the disformal contribution $\mathcal D_{\mu\nu}$, we note that the contraction over the index $\alpha$ generates the `kinetic' term $\chi \equiv -\frac{1}{2}\Phi^\alpha \Phi_\alpha$ leading to
\begin{align}
    F^\alpha _{\mu \alpha }
	&=
	\frac{ 
		nA_{;\mu }
	}{2A }
 	+
	\frac{\Phi^\alpha\Phi _{(\mu;\alpha )}}{A - 2\chi } 
    +
	\frac{
		A _{;\mu }\chi
	}{A (A - 2\chi )} 
	\label{KGFS3}\\[0.5em]
	&\quad
	+\,
	\left[
		\frac{g^{\alpha \beta }}{A}
		-
		\frac{\Phi ^\alpha \Phi ^\beta }{A(A - 2\chi )} 
	\right]
	(
		\Phi _\mu\Phi _{[\beta;\alpha ]} 
		+
		\Phi _{\alpha }\Phi _{[\beta;\mu]} 	
	).
    \nonumber
\end{align}
By virtue of the `orthogonality' condition and the conformal-disformal constraint given by (\ref{Bxdisft}) we find
\begin{align}
    \label{a2chi}
    A - 2\chi = b^2 A^{3 - n}.
\end{align}
In effect, 
\begin{align}
	F^\alpha _{\mu \alpha }
	&=
	\frac{nA_{;\mu }}{2A }
	+
	\frac{\chi A _{;\mu }}{b^2 A^{4-n}} 
	+
	\frac{\Phi^{\alpha  }\Phi _{[\alpha ;\mu]} 	}{A}
	+
	\frac{2\chi \Phi ^\alpha \Phi _{[\alpha;\mu]} }{b^2A^{4-n}} 
    \nonumber
    \\[0.5em]
	&\quad
    +\,
	\frac{\Phi^\alpha\Phi _{(\mu;\alpha )}}{b^2 A^{3-n} }.
\end{align}

Upon expansion of the terms with symmetric and anti-symmetric indices on the right hand side and invoking once again the `orthogonality' condition, we get
\begin{align}
	F^\alpha _{\mu \alpha }
	&=
	\frac{A_{;\mu }}{A}.
\end{align}
This is a remarkable simplification coming from a relatively complicated expression for $F^\alpha_{\mu\nu}$ given by (\ref{Fmunu})! Further observe that even though $F^\alpha_{\mu\nu}$ depends on the number of spacetime dimensions, $F^\alpha_{\mu\alpha}$ and consequently, its covariant derivative, does not. In particular, we have
\begin{align}
    \label{Faman}
	F^\alpha _{\mu \alpha ;\nu }
	&=
	\frac{A_{;\mu \nu }}{A} 
	-
	\frac{A_{;\mu }A_{;\nu }}{A^2}.
\end{align}

The second disformal contribution in (\ref{dcontribrmn}) is not as `forgiving' as $F^\alpha_{\mu\alpha;\nu}$ even with the triple KG constraints. The resulting expression for $F^\alpha_{\mu\nu;\alpha}$ is rather lengthy. Nonetheless, the calculation is  straightforward. We have
\begin{align}
    \label{Famna}
    F^\alpha_{\mu\nu;\alpha}
    &=
    \frac{A_{;\mu \nu }}{A} 
	-
	\frac{A_{;\mu }A_{;\nu }}{A^2} 
	-\bigg[
		\frac{{\Phi ^\alpha }_{;\alpha }}{b^2A^{4-n}}
		+
		\frac{(n-4)A_{;\alpha }\Phi ^\alpha }{b^2 A^{5-n}} 
	\bigg] 
	A_{(;\mu }\Phi _{\nu)} 
	\nonumber
	\\[0.5em]
	&\quad
	+\,
	\bigg[
		\frac{A_{;\beta  }\Phi ^\beta  \Phi ^\alpha _{;\alpha }
		+ 
		A_{;\beta }\Phi ^\alpha \Phi ^\beta _{;\alpha }	
		+
		\Phi ^\alpha\Phi ^\beta A_{;\alpha \beta }  
	}{2b^2A^{4-n}}
    \nonumber
    \\[0.5em]
    &\qquad
    +\,
    \frac{(n-4)A_{;\alpha }A_{;\beta }\Phi ^\alpha \Phi ^\beta}
    {2b^2A^{4-n}}
    +
    \frac{A^{;\alpha }A_{;\alpha } - A\,\square A}{2A^2} 
    \bigg] g_{\mu\nu}
	\nonumber
	\\[0.5em]
	&\quad
	-\,
	\frac{\Phi ^\alpha (
		A_{;\alpha (\mu }\Phi _{\nu)} 
		+
		A_{(;\mu }\Phi _{\nu); \alpha }
	)}{b^2 A^{4-n}}
	\nonumber
	\\[0.5em]
	&\quad
	-\,\bigg\{\frac{A_{;\alpha }}{A}\bigg[
		\frac{g^{\alpha \beta }}{A}
		+
		\frac{(n-4)\Phi ^\alpha \Phi ^\beta }{b^2 A^{4-n}}  
	\bigg] 
	+
	\frac{\Phi^\alpha _{;\alpha }\Phi ^\beta + \Phi ^\alpha \Phi ^\beta _{;\alpha } }{b^2 A^{4-n}} \bigg\}
    \nonumber
    \\[0.5em]
	&\qquad
    \times\,
    (
		\Phi _\mu\Phi _{[\beta;\nu ]} 
		+
		\Phi _{\nu }\Phi _{[\beta;\mu]} 	
	)
    \nonumber
    \\[0.5em]
    &\quad
	+\,
	\bigg[
		\frac{\Phi^\alpha_{;\alpha }}{b^2 A^{3-n} }
		+
		\frac{(n-3)A_{;\alpha}\Phi^\alpha}{b^2 A^{4-n} }
	\bigg]\Phi _{(\mu;\nu )} 
	\nonumber
	\\[0.5em]
	&\quad
	+\,
	\left(
		\frac{g^{\alpha \beta }}{A}
		-
		\frac{\Phi ^\alpha \Phi ^\beta }{b^2A^{4-n}} 
	\right)	
    (\Phi _{\mu;\alpha }\Phi _{[\beta;\nu ]} 
	+
	\Phi _{\nu;\alpha }\Phi _{[\beta;\mu]}) 	
	\nonumber
	\\[0.5em]
	&\quad
	+\,
	\left(
		\frac{g^{\alpha \beta }}{A}
		-
		\frac{\Phi ^\alpha \Phi ^\beta }{b^2A^{4-n}} 
	\right)	
	(\Phi _\mu\Phi _{[\beta;\nu ];\alpha } 
	+
	\Phi _{\nu }\Phi _{[\beta;\mu];\alpha }) 			 
	\nonumber
	\\[0.5em]
	&\quad
	+\,
	\frac{
		\Phi^\alpha\Phi _{(\mu;\nu );\alpha }
	}{b^2 A^{3-n} }.	
\end{align}
Note that the last two lines above contain third-order derivative terms involving $\Phi_\mu$ and ultimately, the field variables $(\phi, X, Y, Z)$. In fact, $F^\alpha_{\mu\nu;\alpha}$ is the only source of these higher order derivative terms in the calculation for $\widehat R_{\mu\nu}$, coming from the covariant derivative of $F^\alpha_{\mu\nu}$. That is, given the set of KG constraints without which the term $F^\alpha_{\mu\alpha;\mu}$ could have added more third order derivative terms. With this being said, at first glance, these terms may look quite problematic, at least if we consider working on the field variables $(\phi, X, Y, Z)$. However, because the third-order derivative factors are multiplied with at most first-order derivative factors,\footnote{e.g., $\Phi_{[\beta;\nu];\alpha}$ is multiplied with $\Phi_\mu$ and functional of $A$.} then we may simply perform integration by parts and effectively transfer one of the derivative indices in the third-order derivative terms to its `neighbours'. We deal with these terms in the next section upon derivation of the disformal Ricci scalar.

The last two disformal contributions in (\ref{dcontribrmn}) are simply contractions of two $F^\alpha_{\mu\nu}$'s and a product of $F^\alpha_{\rho\alpha}$, already calculated above, and $F^\rho_{\mu\nu}$. Upon adding these two terms to the sub-results (\ref{Faman}) and (\ref{Famna}), and combining similar terms in the equation for $\widehat R_{\mu\nu} $ given by (\ref{hattedRicci}) above, we find
\begin{align}
    \label{hatRmn}
	&
	\widehat R_{\mu \nu } - R_{\mu \nu }
	\nonumber
	\\[0.5em]
	&=
	\bigg[ 
		\bigg(
			\frac{n-5}{A^3} 
			-
			\frac{1}{b^2 A^{5-n}}
		\bigg) (n-3)\frac{\Phi ^\alpha A_{;\alpha } }{2}
        \nonumber
        \\[0.5em]
        &\qquad
        +\,
        \bigg(
			\frac{n-3}{A^2} 
			-
			\frac{1}{b^2A^{4-n}}
		\bigg)\frac{{\Phi ^\alpha }_{;\alpha }}{2} 
	\bigg] 
	A_{(;\mu }\Phi _{\nu)} 
	\nonumber
	\\[0.5em]
	&\quad
	+\,
    \bigg[
        \frac{A^{;\alpha }A_{;\alpha } - A\,\square A}{2A^2}
    	+
    	\frac{A_{;\beta  }\Phi ^\beta  \Phi ^\alpha _{;\alpha }
    	+ 
    	A_{;\beta }\Phi ^\alpha \Phi ^\beta _{;\alpha }	
    	}{2b^2A^{4-n}}
        \nonumber
        \\[0.5em]
        &\qquad\quad
        +\,
        \frac{
        \Phi ^\alpha\Phi ^\beta A_{;\alpha \beta }  
        +
        (n-4)A_{;\alpha }A_{;\beta }\Phi ^\alpha \Phi ^\beta
    	}{2b^2A^{4-n}} 
    \bigg]g_{\mu\nu}
    \nonumber
    \\[0.5em]
    &\quad
    -\,
    \frac{\Phi ^\alpha (
        A_{;\alpha (\mu }\Phi _{\nu)} 
        +
        A_{(;\mu }\Phi _{\nu); \alpha }
    )}{b^2 A^{4-n}}
	\nonumber
	\\[0.5em]
	&\quad
	+\,
	\left(
		\frac{g^{\alpha \beta }}{A}
		-
		\frac{\Phi ^\alpha \Phi ^\beta }{b^2A^{4-n}} 
	\right)
    (\Phi _{\mu;\alpha }\Phi _{[\beta;\nu ]} 
    +
    \Phi _{\nu;\alpha }\Phi _{[\beta;\mu]})	
    \nonumber
    \\[0.5em]
    &\quad
	+\,
	\left(
		\frac{g^{\alpha \beta }}{A}
		-
		\frac{\Phi ^\alpha \Phi ^\beta }{b^2A^{4-n}} 
	\right)	
    (\Phi _\mu\Phi _{[\beta;\nu ];\alpha } 
    +
    \Phi _{\nu }\Phi _{[\beta;\mu];\alpha })   
	\nonumber
	\\[0.5em]
	&\quad
	+\,
	\bigg[
		\frac{\Phi^\alpha_{;\alpha }}{b^2 A^{3-n} }
		+
		\frac{(n-4)A_{;\alpha}\Phi^\alpha}{b^2 A^{4-n} }
	\bigg]\Phi _{(\mu;\nu )}
	+
	\frac{
		\Phi^\alpha\Phi _{(\mu;\nu );\alpha }
	}{b^2 A^{3-n} }
    \nonumber
    \\[0.5em]
    &\quad
	-\,
	\frac{\Phi ^\alpha \Phi ^\beta }{2b^2A^{4-n}} 
	\Phi _{\mu ;\alpha }\Phi _{\nu ;\beta }
 	-
	\bigg(
		\frac{2A^{;\alpha }}{A^2}
		+
		\frac{{\Phi ^\alpha }_{;\beta }\Phi ^\beta }{b^2A^{4-n}}  
	\bigg)
	\Phi _{\alpha (;\mu }\Phi _{\nu )}	 
	\nonumber
	\\[0.5em]
	&\quad
	-\,
	\frac{1}{4} \bigg[\frac{(n-3)^2}{A^2} 
		- 
		\frac{b^2(n-3)^2}{2A^n} 
		-
		\frac{1}{2b^2A^{4-n}} 
	\bigg]A_{;\mu }A_{;\nu }
	\nonumber
	\\[0.5em]
	&\quad
	-\,
	\bigg\{
		\bigg[
			\frac{b^2(n-3)^2}{2A^{n+1}} 
			-
			\frac{n-3}{A^3} 
			+
			\frac{1}{2b^2A^{5 - n}} 
		\bigg]\frac{A^{;\alpha }A_{;\alpha }}{4} 
        \nonumber
        \\[0.5em]
        &\qquad\quad
		+\,
		\bigg(
			\frac{n-3}{A^3} 
			-
			\frac{1}{b^2 A^{5 - n}} 
		\bigg)\frac{A^{;\alpha }\Phi _{\alpha ;\beta }\Phi ^{\beta }}{2}
		\nonumber
		\\[0.5em]
		&\qquad\quad
		-\,
		\frac{\Phi ^{\alpha ;\beta }\Phi _{[\alpha ;\beta ]}}{A^2} 
		+
		\frac{\Phi ^{\alpha ;\rho }\Phi _\rho 
			\Phi _{\alpha ;\beta }\Phi ^\beta 
		}{2b^2A^{5 - n}} 		
	\bigg\}\Phi _\mu \Phi _\nu
	\nonumber
	\\[0.5em]
	&\quad
	-\,
	\frac{1}{2}\bigg(
		\frac{1}{b^2 A^{3-n}} - \frac{1}{A} 
	\bigg) 
	(
	{\Phi _\mu }^{;\alpha }{\Phi _{\nu ;\alpha }}
	-
	{\Phi ^\alpha  }_{;\mu  }{\Phi _{\alpha  ;\nu  }}
	)
	\nonumber
	\\[0.5em]
	&\quad
	+\,
	\bigg[
		\bigg(
			\frac{n-5}{A} \Phi ^\beta  A_{;\beta  }
			+
			{\Phi ^\beta  }_{;\beta  }
		\bigg)\frac{\Phi ^\alpha }{b^2A^{4-n}} 
        \nonumber
        \\[0.5em]
        &\qquad\quad
		+\,
		\bigg(
			\frac{1}{b^2A^{4-n}} 
			-
			\frac{n-7}{A^2} 
		\bigg)\frac{A^{;\alpha }}{2} 
	\bigg]\Phi _{(\mu }\Phi _{\nu) ;\alpha }.
\end{align}
The disformal contributions on the right hand side depend on the conformal and disformal factors $(A, C, D)$, and their covariant derivatives which ultimately act on the field variables $(\phi, X, Y, Z)$. Due to the `orthogonality' condition and the conformal-disformal constraint, both disformal factors do not have independent existence of their own; they are both expressible in terms of the conformal factor and $(\phi, X, Y, Z)$. Each term in the equation above for $\widehat R_{\mu\nu} - R_{\mu\nu}$ is manifestly symmetric with respect to the indices $(\mu,\nu)$ reflective of the same for $\widehat R_{\mu\nu}$ and $R_{\mu\nu}$. In addition to this, we have dependence on the spacetime dimension $n$ which adds another layer of complication in the expression for the disformal Ricci tensor.

As a consistency check for the result above for $\widehat R_{\mu\nu}$, we take its limit as the conformal factor $A\rightarrow 1$. By virtue of conformal-disformal constraint given by (\ref{Bxdisft}), the disformal factor $C$ correspondingly vanishes. Due to the `orthogonality' condition binding the disformal factors, this implies the other disformal factor $D\rightarrow 0$ as well. Consequently, we recover the original Ricci tensor from (\ref{hatRmn}); that is, $\widehat R_{\mu\nu} \rightarrow R_{\mu\nu}$, as expected. Note that following this line of thinking, we cannot naively take the limit of $\widehat R_{\mu\nu}$ given above by taking $\Phi_\mu \rightarrow 0$ alone. Due to the triple KG constraints imposed on $\widehat R_{\mu\nu}$ (see the last part of Sec. \ref{seckgdisf}), the limit $\Phi_\mu \rightarrow 0$ is tantamount to $A,C,D\rightarrow 0$, which gives a trivial result. 

Although relatively lengthy, the constrained disformally transformed Ricci tensor given by (\ref{hatRmn}) is still much simpler compared to its unconstrained counterpart given in Appendix A; see (\ref{hatRmnNocons}). The latter is at least twice as long as the former. Moreover, consistent with the fact that (\ref{hatRmn}) is a special case of (\ref{hatRmnNocons}), the unconstrained one has extra third-order and lower-order derivative terms not present in the constrained relation; e.g., those terms involving $\chi$ and derivative thereof that could have been simplified through the `orthogonality' condition. In the following section, we see further significant simplification with the derivation of the disformally transformed action. Naively, the contraction of the hatted inverse metric with the hatted Ricci tensor above could produce a longer and more complicated result. But with the triple KG constraints and `removal' of third-order derivative terms, we find a much shorter six-term disformal contributions in the disformally transformed Einstein-Hilbert action in $n$ dimensions.

\section{Klein-Gordon constrained disformal transformation of the Einstein-Hilbert action}
\label{secdtehact}

\noindent 
In this section, we aim to express the disformally transformed Einstein-Hilbert action in terms of the original Ricci scalar and terms contributed by the KG-constrained disformal transformation. The hatted action, $\widehat S$, borrows its form from the original one given by (\ref{ehaction}). We have in $n$ spacetime dimensions,
\begin{align}
    \widehat S 
    = 
    \frac{1}{2}
    \int \dd^n x\,\sqrt{-\widehat g}\, \widehat R,
\end{align}
where $\widehat g$ and $\widehat R$ are the disformally transformed metric determinant and Ricci scalar, respectively. In spite of the simple and concise expression for the Einstein-Hilbert action, due to $R$ being rooted in the metric and its derivatives, disformal transformation can produce a lengthy and complicated equation for $\widehat S$. Our hope is that with the help of the triple KG constraints in Sec. \ref{seckgdisf}, we can somehow mend the resulting action leading to a much simpler form, and identify those beyond-Horndeski terms that will call for deeper meanings and implications in our future studies.

On the one hand, we have the hatted integral measure, $\dd^nx\,\sqrt{-\widehat g}$. The hatted metric determinant, $\widehat g$, in this measure can be computed following the matrix determinant lemma applied to the general disformal transformation (\ref{genDT}). Identifying $\{\Phi_\mu\Phi_\nu\}$ in this transformation as a dyadic product added to an invertible square matrix, $\{Ag_{\mu\nu}\}$, we find
\begin{align}
    \widehat g 
    = 
    \det\{Ag_{\mu\nu} + \Phi_\mu\Phi_\nu\}
    =
    A^{n-1}(A - 2\chi)g\label{gdetGDT}
\end{align}
By virtue of (\ref{a2chi}), this simplifies to $\widehat g = b^2 A^2 g$. In effect, the integral measure in the equation for $\widehat S$ becomes
\begin{align}
    \dd^n x \sqrt{-\widehat g} 
    = 
    bA \,\dd^n x \sqrt{-g}.
\end{align}
Observe that the factor, $bA$, multiplied to the original integral measure does not depend on the number of spacetime dimensions. The `orthogonality' condition removed the dependence on $n$ of the Jacobian multiplying the integral measure $\dd^n x\sqrt{-g}$.

On the other hand, we have the disformally transformed Ricci scalar whose calculation start with the immediately preceding section involving the hatted Ricci tensor; that is, $\widehat R = \widehat g^{\mu\nu}\widehat R_{\mu\nu}$. By virtue of (\ref{invg}) and (\ref{hatRmn}) for the inverse disformal metric and the hatted Ricci tensor, respectively, we find 
\begin{align}
    \label{Rhat}
    \widehat R
    =
    \frac{R}{A}
    -
    \frac{R_{\mu\nu}\Phi^\mu\Phi^\nu}
    {b^2 A^{4-n}}
    +
    \widehat g^{\mu\nu}\mathcal D_{\mu\nu},
\end{align}
where the disformal contribution in $\widehat R_{\mu\nu}$, namely $\mathcal D_{\mu\nu}$, is as defined by (\ref{dcontribrmn}). In other words, the disformally transformed Ricci scalar is a sum of the original Ricci scalar modulo the conformal factor and a set of disformal contributions---now with the original Ricci tensor---encoded in the last two terms above.

Substitution of the equations above for $\widehat R$ and the integral measure in equation for the hatted action yields
\begin{align}
    \widehat S
    &=
    \frac{1}{2}\int
    \dd^n x\,\sqrt{-g}\bigg(
        bR 
        -
        \frac{A^{n-3}}{b}
        R_{\mu\nu}\Phi^\mu\Phi^\nu
        +
        bA\widehat g^{\mu\nu}\mathcal D_{\mu\nu}
    \bigg).
    \nonumber
\end{align}
In Sec. \ref{seckgdisf}, we find third derivative terms in $\mathcal D_{\mu\nu}$ in the field variables $(\phi, X, Y, Z)$; specifically, these are
\begin{align}
	\left(
		\frac{g^{\alpha \beta }}{A}
		-
		\frac{\Phi ^\alpha \Phi ^\beta }{b^2A^{4-n}} 
	\right)	
    (\Phi _\mu\Phi _{[\beta;\nu ];\alpha } 
    +
    \Phi _{\nu }\Phi _{[\beta;\mu];\alpha })  
    +
    \frac{
        \Phi^\alpha\Phi _{(\mu;\nu );\alpha }
	}{b^2 A^{3-n} }
    \nonumber
\end{align}
Now that we have the hatted action in hand, we can `remove' these higher order derivative terms by performing integration by parts. For the terms above, this results in
\begin{align}
    &\frac{\Phi^{\mu;\alpha}\Phi_\alpha
        \Phi_{\mu;\nu}\Phi^\nu}
        {b^4A^{7-2n}}
    +
    \frac{\Phi^{\mu;\nu}
        (\Phi_{\mu;\nu} - \Phi_{\nu;\mu})}
        {b^2A^{4-n}}
    -
    \frac{{\Phi^\mu}_{;\mu}{\Phi^\nu}_{;\nu}}
        {b^2A^{4-n}}
    \nonumber
    \\[0.5em]
    &-\,
    \bigg[
        \frac{3(n-3)}{A^{5-n}}
        +
        \frac{1}{b^2 A^{7-2n}}
    \bigg]\frac{
        A^{;\mu}\Phi_{\mu;\nu}\Phi^\mu
        - 
        \Phi^\mu A_{;\mu} {\Phi^\nu}_{;\nu}}{2b^2}
    \nonumber
    \\[0.5em]
    &-\,
    \frac{(n-3)A^{;\mu}A_{;\mu}}{2}
    \bigg(
        \frac{n-3}{A^3}
        +
        \frac{1}{b^2 A^{5-n}}
    \bigg)
    \nonumber
    \\[0.5em]
    &-\,
    \bigg(
        \frac{n-3}{A^{6-n}}
        +
        \frac{1}{b^2 A^{8 - 2n}}
    \bigg)
    \frac{(n-3)\Phi^\mu A_{;\mu}\Phi^\nu A_{;\nu}}{b^2}.    
\end{align}

Using this expression in $\mathcal D_{\mu\nu}$ and 
performing additional series of integrations by parts to simplify the resulting action, we finally obtain the desired action in general spacetime dimension $n$.
\begin{align}
    \label{dtehs}
    \boxed{
    \widehat S
    =
    \frac{1}{2}\int \dd^nx\,\sqrt{-g}\,\bigg(
        bR 
        - 
        \frac{A^{n-3}}{b}
        \Phi^\mu\Phi^\nu R_{\mu\nu}
        +
        bA\tilde {\mathcal D}
    \bigg),
    }
\end{align}
where
\begin{align}
    \label{tildeD}
    \tilde {\mathcal D}
    &=
    \frac{\Phi^{\mu;\alpha} 
        \Phi_{\mu;\nu} \Phi^{\nu } \Phi_{\alpha}}
        {2b^2A^{5-n}}
    +
    \bigg(  
        \frac{1}{b^2A^{4-n}}
        -
        \frac{1}{A^2}
    \bigg)
    \Phi^{\mu;\nu}\Phi_{[\mu;\nu]} 
    \nonumber
    \\[0.5em]
    &\quad
    +\,
    \bigg(
        \frac{n-3}{2 {A^{3}}}
        -
        \frac{2n-5}{2 b^2A^{5-n}}
    \bigg) 
    A^{;\mu}\Phi_{\mu;\nu }\Phi^{\nu}
    \nonumber
    \\[0.5em]
    &\quad
    -\,
    \frac{(n-3)( n-2)}{4b^2A^{6-n}}
    \Phi^\mu A_{;\mu}\Phi^\nu A_{;\nu}
    \\[0.5em]\label{KGRicScalD}
    &\quad    
    +\,
    \left[\frac{b^2(n-3)^2}{8A^{n+1}}
    -
    \frac{( n-3)^2}{4A^3}
    +
    \frac{1}{8 b^2A^{5-n}}\right]
    A^{;\mu}A_{;\mu}
    \nonumber 
\end{align}
Such a disformally transformed action should be compared to the unconstrained hatted action shown in Appendix B. The triple KG constraints specified in Sec. \ref{seckgdisf} have given us a much shorter and hence, simpler transformed Einstein-Hilbert action that is at most second order in the field variables $(\phi, X, Y, Z)$.

Having said this, a keen reader may wonder if the hatted action above can be further simplified, say, by way of integration by parts or constructing Ricci or curvature tensor from the disformal contributions. On the one hand, knowing the existence of the Ricci tensor in form of $R_{\mu\nu}\Phi^\mu\Phi^\nu$, it might be tempting to construct the Riemann curvature tensor out of the term involving $\Phi^{\mu;\alpha}\Phi_{\mu;\nu} \Phi^{\nu } \Phi_{\alpha}$ on the right hand side of the equation above for $\tilde {\mathcal D}$; that is, with the intention to cancel out $R_{\mu\nu}\Phi^\mu\Phi^\nu$ by `transferring' the covariant derivative index $\nu$ from $\Phi_{\mu;\nu}$ to $\Phi^{\mu;\alpha}$. However, such an attempt leads to 
\begin{align}
    &\Phi^{\mu;\alpha}\Phi_{\mu;\nu} \Phi^{\nu } \Phi_{\alpha}
    =
    (-2\chi^{;\alpha})_{;\nu} \Phi^\nu \Phi_{\alpha}
    -
    \Phi_{\mu;\alpha \nu}\Phi^\mu \Phi^\nu \Phi^\alpha.
    \nonumber
\end{align}
The second term on the right hand side has $\Phi$ with three indices but they are made symmetric by the factor $\Phi^\mu\Phi^\nu\Phi^\alpha$. We cannot certainly construct a Riemann curvature tensor and consequently, the Ricci tensor out of this. In fact, `forcefully' generating the Riemann tensor from (\ref{tildeD}) and Ricci tensor coming from its index contraction, by adding or subtracting third order derivative terms, may only create lengthier expressions. 

On the other hand, one may be compelled to `spell out' the Ricci tensor in (\ref{dtehs}) instead of creating one out of $\Phi_\mu$'s to cancel it. This seems a plausible way of reducing the number of terms in the equation for $\widehat S$ because
\begin{align}
    \Phi^\mu \Phi^\nu R_{\mu\nu}
    =
    \Phi^\mu(
        {\Phi^\alpha}_{;\mu\alpha}
        -
        {\Phi^\alpha}_{;\alpha\mu}
    ).
\end{align}
However, on performing integration by parts in an attempt to remove the third-order derivative terms in the hatted action, we find other terms previously removed by integration by parts in the process of arriving at (\ref{dtehs}) to be `regenerated'; e.g., ${\Phi^\alpha}_{;\alpha}{\Phi^\beta}_{;\beta}$. This then leads to a  longer expression. As such, we find it prudent, at least in this work, to stick to the form of the KG-constrained disformally transformed action given by (\ref{dtehs}). We leave it for future work the other use cases requiring further manipulations of this equation, be they resulting to a lengthier expression, for exploration of their deeper meanings and possible applications within the framework of scalar tensor theory.

Before we leave this section and explore in the next section the special cases of the resulting action, it is worth reflecting at this point about the physical degrees of freedom involved in the current work, in relation to a related studiy mentioned above involving the disformal transformation of the action. In Ref. \cite{Bettoni:2013diz}, the authors started with the Horndeski action itself and perform special disformal transformation. Knowing that this transformation is invertible and that the resulting form is the same as the original one, our insight is that there is effectively no new physical degrees of freedom `generated'. Within the framework of inflationary cosmology for instance, such transformation leaves the scalar and tensor primordial cosmological perturbations invariant at least in the superhorizon limit; see for instance, Refs. \cite{Tsujikawa:2014uza,Domenech:2015hka,Motohashi:2015pra,Alinea:2020laa,Minamitsuji:2014waa}. 

In the current work however, the original action is simply a special case of the Horndeski action, namely, the Einstein-Hilbert action. In the absence of additional terms in the original action that could possibly lead to some cancellation and owing to the generality and complexity of the disformal transformation, we gain a more general action that includes two sub-Lagrangians of the Horndeski action and several beyond-Horndeski terms. These new Horndeski and beyond-Horndeski terms are suggestive of new physical degrees of freedom rooted in the conformal and disformal factors. However, we have the insight that as the disformal factors are linked to the `free' conformal factor by the set of KG constraints, that is, both disformal factors are expressible in terms of the conformal factor, we may only have at most one additional physical degree of freedom. This is in addition to the fact that from the perspective of single scalar field Scalar-Tensor theory, we have ultimately introduced only one new field, namely, $\phi$. Beyond this insight, as to the nature and full accounting of this possible new physical degree of freedom, we see further needs for more thorough analyses and in-depth calculations. We hope to address this matter in our future studies.

\section{The KG-constrained disformally transformed Einstein-Hilbert action in \boldmath\texorpdfstring{$n = 4, 3, 2$}{n = 4, 3, 2} dimensions}
\label{secDimlim}

Content with the form of the KG-constrained disformally transformed Einstein-Hilbert action in general spacetime dimensions, in this section, we explore the special cases for $n = 4, 3, 2$ spacetime dimensions. The consideration for the special case involving $n = 4$ warrants no justification. We could have, in fact, simply set our goal to derive the disformally transformed action in four dimensions right from the very beginning. The Einstein-Hilbert action was first formulated in four spacetime dimensions \cite{Hilbert:1915tx}. And the gamut of its successes, past more than a century of its existence, are for the most part, still within the framework of four dimensions; see for instance, Refs. \cite{Ishak:2018his,Robertson:1933zz}. The macroscopic world where we live is four dimensional. It is then natural that we look into any extensions of the Einstein-Hilbert action in four dimensions. 

Having said this, the cases for lower dimensions may find significance in their much simpler forms. In the realm of sub-atomic world, theories in lower dimensions may provide insights into the workings of nature free from the added complications of extra dimensions. There are theories for instance, within the framework of quantum field theory, dealing with lower dimensions that can elucidate the behaviour of a physical system with much simpler mathematics involved but can lead to important consistency relations, links to higher dimensions, and limits that may pave the way for some experimental tests aiming to check the veracity of one whole `big theory' for some special systems. While the main value that we practically prioritise in this work is the aforementioned simplicity of the disformally transformed Einstein-Hilbert action in lower dimensions, we hope that some further connections in the quantum realm be made in future studies.

\par
\vspace{1.0em}
\begin{center}
\begin{tabular}{ c c c c c}
\toprule
disformal \\ contribution \\ terms & $n$ & $n = 4$ & $n = 3$ & $n = 2$\\
\midrule
\\[0.5em]
$\Phi ^\mu \Phi ^\nu R_{\mu \nu }$ & $\checkmark$ & $\checkmark$ & $\checkmark$ & $\bigcirc$
\\[0.5em] 
$\Phi^{\mu ;\alpha }\Phi_{\mu ;\nu} \Phi^\nu \Phi_\alpha$ & $\checkmark$ & $\checkmark$ & $\checkmark$ & $\bigcirc$
\\[0.5em] 
$\Phi^{\mu ;\nu }\Phi_{[\mu ;\nu]}$ & $\checkmark$ & $\checkmark$ & $\checkmark$ & $\bigcirc$
\\[0.5em] 
$A^{;\mu }\Phi_{\mu ;\nu }\Phi^\nu$ & $\checkmark$ & $\checkmark$ & $\checkmark$ & $\bigcirc$
\\[0.5em] 
$\Phi^\mu A_{;\mu }\Phi^\nu A_{;\nu }$ & $\checkmark$ & $\checkmark$ & $\bigcirc$ & $\bigcirc$
\\[0.5em]
$A^{;\mu}A_{;\mu}$ & $\checkmark$ & $\checkmark$ & $\checkmark$ & $\bigcirc$
\\
\bottomrule
\end{tabular}
\captionof{table}{\small Disformal contribution terms in the KG-constrained disformal transformation of the Einstein-Hilbert action, in general $n$ dimensions and in $n = 4, 3, 2$ dimensions. The check mark (circle) $\checkmark\,(\bigcirc)$ indicates the presence (absence) of a given term.}
\end{center}
\par
\vspace{1.0em}

\underline{Case $n = 4$ dim.} There are five disformal contribution terms in addition to that involving the Ricci tensor inside the pair of parentheses in (\ref{dtehs}); see also Table 1. In four spacetime dimensions we still have six disformal contribution terms. Setting the constant $b = 1$ we obtain\footnote{One may wonder if the constant $b$ can be set other than unity to yield a much shorter expression. However, upon short inspection of (\ref{dtehs}) and (\ref{tildeD}) for the action, we find that $b$ appears as always `coupled' with $A$ in expressions involving sums of terms containing $A$; e.g., $(n-3)/2A^3 - (2n - 5)/2b^2A^{5-n}$. With $A$ being dimensionless, no adjustment of $b$ can zero out expressions involving sums involving different powers of $A$. For instance, the expression $(n-3)/2A^3 - (2n - 5)/2b^2A^{5-n}$ in four dimensions becomes $1/2A^3 - 4/2b^2A$, and no adjustment of $b$ can make it vanish.}
\begin{align}
    \label{S4act}
	\widehat S^{(4)}
	&=
	\frac{1}{2}\int \dd^4 x \sqrt{-g}\, \bigg[
		R
		-
		A\Phi ^\mu \Phi ^\nu R_{\mu \nu }
		+
  \frac{1}{2} \Phi^{\mu ;\alpha }\Phi_{\mu ;\nu} \Phi^\nu \Phi_\alpha 
		\nonumber
		\\[0.5em]
		&\quad\quad	
		+\,
		\frac{A^2 - 1}{A}
		\Phi^{\mu ;\nu }\Phi_{[\mu ;\nu]}  
		+
		\frac{1 - 3A^2}{2A^2}A^{;\mu }\Phi_{\mu ;\nu }\Phi^\nu
		\nonumber
		\\[0.5em]
		&\quad\quad  
		-\,
		\frac{\Phi^\mu A_{;\mu }\Phi^\nu A_{;\nu }}{2A}
		+
		\frac{(A^2 - 1)^2}{8A^3}A^{;\mu }A_{;\mu }  
	\bigg].
\end{align}
We have a rather concise expression above for the disformally transformed action compared to the unconstrained action given by (\ref{unconstrhatS}) supplemented by (\ref{tildeH}), for $n = 4$, given in Appendix B. 

While acknowledging a significant simplification of the transformed action due to the triple KG constraints, we also recognise the existence of numerous terms involving contractions and products of expressions that ultimately depend on $(\phi,X,Y,Z)$, hiding in the disformal contribution in the hatted action $S^{(4)}$ above. The simplest of them involves 
\begin{align}
    \label{AmuAmu}
    A_{;\mu}A^{;\mu}
    =
    -2XA_{\phi}^2 + 2YA_\phi A_X + A_X^2 Z,
\end{align}
where $A_\phi = \partial_\phi A$ and $A_X = \partial_\phi A$. The other terms in the disformal contribution can be expanded in terms of $A$ and $(\phi,X,Y,Z)$ and their derivatives by virtue of the `orthogonality' condition (\ref{orthocond}) yielding $2CX = DY$ and the conformal-disformal constraint (\ref{cdconstr}) relating $C^2$ with $A$ and $(X,Y,Z)$. The full expression for this expansion is too lengthy to be written out explicitly in this paper but it is good to be aware of their existence.

With this being said, one may wonder if in the `forest' of terms in the full expansion of $S^{(4)}$ we may find fragments of the Horndeski action \cite{Horndeski:1974wa}---the most general gravitational action within the framework of Scalar Tensor theory in four dimensions involving the metric tensor and a scalar field, whose equations of motion are at most second order. It reads
\begin{align}
    S_H
    &=
    \int \sqrt{-g}\,\dd^4x\,(\mathcal L_2 + \mathcal L_3 + \mathcal L_4 + \mathcal L_5),
\end{align}
where
\begin{align}
    \mathcal L_2 
    &= 
    G_2,
    \nonumber
    \\[0.5em]
    \mathcal L_3 
    &= 
    G_3\,\square \phi,
    \nonumber
    \\[0.5em]
    \mathcal L_4 
    &= 
    G_4 R + G_{4,X}[(\square\phi)^2 - \phi_{;\mu\nu}\phi^{;\mu\nu}],
    \quad (G_{4,X} = \partial_X G_4)
    \nonumber
    \\[0.5em]
    \mathcal L_5
    &=
    -
    \frac{1}{3!}G_{5,X}[
        (\square\phi)^3 
        + 
        2\phi^{;\mu\alpha}\phi_{;\alpha\beta}{\phi^{;\beta}}_\mu
        -
        3\phi^{;\mu\nu}\phi_{;\mu\nu}\,\square\phi
    ]
    \nonumber
    \\[0.5em]
    &\quad
    +\,
    G_5 G_{\mu\nu}\phi^{;\mu\nu}, 
\end{align}
where $G_2$ to $G_5$ are all functionals of $(\phi,X)$ and $G_{\mu\nu} = R_{\mu\nu} - \frac{1}{2}g_{\mu\nu}R$ is the Einstein tensor\footnote{It encodes many gravitational theories including General Relativity and Brans-Dicke theory, to name just a couple of them. Indeed, when $ G_4 \sim \text{const.} $ and all other $G_i$'s are vanishing, we get the (geometric part) of the Einstein-Hilbert action. On the other hand, if $ G_4\sim \phi$ and $G_2 \sim X/\phi$ with $ G_3, G_5 = 0$, we get the Brans-Dicke theory\cite{BransDicke}.}. It may be recalled that the \textit{special} disformal transformation of the Einstein-Hilbert action yields a special case of the Horndeski action covering the sub-Lagrangians $\mathcal L_2, \mathcal L_3,$ and $\mathcal L_4$ \cite{Alinea:2020sei}, consistent with the hierarchical propagation of terms observed in Ref. \cite{Bettoni:2013diz}. For our case here, we can also identify terms belonging to the Horndeski sub-Lagrangians but only for $\mathcal L_2$ and $\mathcal L_3$ with all other numerous terms in the action being beyond-Horndeski terms.

Terms belonging to $\mathcal L_2$ can be found in the action $S^{(4)}$ given by (\ref{S4act}) in the form of expressions involving products of $A, A_\phi, $ and $A_X$ and powers of $X$; e.g., $-2XA_\phi^2$ in $A_{;\mu}A^{;\mu}$ given by (\ref{AmuAmu}). Terms involving $\square \phi$ in the third sub-Lagrangian $\mathcal L_3$ can be generated by performing integration by parts involving expressions with $Y = \phi^{;\mu}X_{;\mu}$ multiplied to any factor that is a functional of $(\phi,X)$, say, $H_d(\phi,X)$; e.g., terms $2YA_\phi A_X$ and $A_\phi A Y/2X$ coming from $A^{;\mu}A_{;\mu}$ and $A^{;\mu}\Phi_{;\mu\nu}\Phi^\nu$, respectively. Indeed, for terms of the form $H_d(\phi,X)\phi^{;\mu}X_{;\mu}$ one may take 
\begin{align}
    H(\phi,X)
    \equiv
    \int^X \dd \bar X H_d(\phi,\bar X),
\end{align}
so that $ H_d\phi^{;\mu}X_{;\mu} = H_X \phi^{;\mu}X_{;\mu} = H_{;\mu}\phi^{;\mu} + 2XH_{\phi},$ which upon integration by parts of $H_{;\mu}\phi^{;\mu}$, yields $-H(\phi,X)\,\square \phi + 2X H_{\phi}(\phi,X)$. In other words, terms of the form $H_d(\phi,X)\phi^{;\mu}X_{;\mu}$ can be be effectively decomposed to part $\mathcal L_2$ and part $\mathcal L_3$ of the Horndeski action.

Terms belonging to $\mathcal L_4$ and $\mathcal L_5$ may not be generated from the disformal transformation (\ref{genDT}) of the Einstein-Hilbert action. The triple KG constraints leave us with a lone Ricci scalar in $\widehat S^{(4)}$ coupled with just unity inside the square brackets in (\ref{S4act}). This is different from the case of \textit{special} disformal transformation in Ref. \cite{Alinea:2020sei} where $R$ is coupled with a functional of $(\phi,X)$ whose derivative with respect to $X$ matches the functional factor multiplied to $(\square\phi)^2 - \phi^{;\mu\nu}\phi_{;\mu\nu}$; that is, properly belonging to $\mathcal L_4$. For the current case, we may identify (or `generate' by way of integration by parts) expressions involving\footnote{The expression $(\square\phi)^2 - \phi^{;\mu\nu}\phi_{;\mu\nu}$, for instance, can be `generated' from $\Phi^\mu\Phi^\nu R_{\mu\nu}$.} $(\square\phi)^2 - \phi^{;\mu\nu}\phi_{;\mu\nu}$ and $ (\square\phi)^3 + 2\phi^{;\mu\alpha}\phi_{;\alpha\beta}{\phi^{;\beta}}_\mu - 3\phi^{;\mu\nu}\phi_{;\mu\nu}\,\square\phi$ in $\mathcal L_4$ and $\mathcal L_5$, respectively. However, with a lone Ricci scalar in $\widehat S^{(4)}$, the factors multiplying these expressions may not fit with the definitions of fourth and fifth Horndeski sub-Lagrangians requiring the Einstein tensor and the coefficient functional of $R$ being dependent on $(\phi,X)$.

\underline{Case $n = 3$ dim.} The case for $n = 3$ spacetime dimensions removes one term present in $S^{(4)}$, namely, $\Phi^\mu A_{;\mu}\Phi^\nu A_{;\nu}$; see Table 1. The corresponding action is given by
\begin{align}
	\widehat S^{(3)}
	&=
	\frac{1}{2}\int \dd^3 x \sqrt{-g}\, \bigg (
		R
		-
		\Phi ^\mu \Phi ^\nu R_{\mu \nu }
        +
        \frac{\Phi^{\mu;\alpha} 
            \Phi_{\mu;\nu} \Phi^{\nu } \Phi_{\alpha}}
            {2A}
        \nonumber
        \\[0.5em]
        &\quad        
        +\,
        \frac{A - 1}{A}
        \Phi^{\mu;\nu}\Phi_{[\mu;\nu]} 
        -
        \frac{3A^{;\mu}\Phi_{\mu;\nu }\Phi^{\nu}}{2A} 
        +
        \frac{A^{;\mu}A_{;\mu}}{8A}
	\bigg).
\end{align}
In addition to this, the coefficients of the remaining disformal contribution terms listed in Table 1 are in general, simpler compared to those in $S^{(4)}$; for instance, the coefficient of $A$ for $\Phi^\mu\Phi^\nu R_{\mu\nu}$ in $S^{(4)}$ is replaced by unity in $S^{(3)}$. In fact, only the disformal contribution term, $\Phi^{\mu ;\alpha }\Phi_{\mu ;\nu} \Phi^\nu \Phi_\alpha$, gained a less simple functional factor in going from $n = 4$ to $n = 3$; i.e., from a constant factor of $1/2$ to $1/2A$. Such a general simplification is due to vanishing of sub-terms involving factors of $(n - 3)$ in the equation above for $\tilde {\mathcal D}$ given by (\ref{tildeD}).

\underline{Case $n = 2$ dim.} If we take $n = 2$ in the general action given by (\ref{dtehs}) supplemented by (\ref{tildeD}), four terms are immediately removed from the disformal contribution terms, leaving us with $\Phi ^\mu \Phi ^\nu R_{\mu \nu }$ and $\Phi^{\mu ;\alpha }\Phi_{\mu ;\nu} \Phi^\nu \Phi_\alpha$ in Table 1. The resulting action is given by
\begin{align}
    \widehat S^{(2)}
	=
	\frac{1}{2}\int \dd^2 x\sqrt{-g} \bigg(
		R 
		- 
		\frac{\Phi ^\mu \Phi ^\nu R_{\mu \nu }}{A} 
		+
		\frac{\Phi^{\mu ;\alpha }\Phi_{\mu ;\nu} \Phi^\nu \Phi_\alpha }{2A^2} 
	\bigg).
\end{align}
In addition to this, the conformal-disformal constraint given by (\ref{cdconstr}) tells us that $C^2 = B = 0$ in two dimensions. And because the disformal factors $C$ and $D$ are bound by the `orthogonality' condition (\ref{orthocond}), this means $D = 0$ as well. Consequently, in two dimensions the disformal transformation (\ref{genDT}) becomes a conformal transformation and the Einstein-Hilbert action, under this transformation, remains the same; that is, $\widehat S^{(2)} = S^{(2)}$! 

To further clarify this point, we note that under conformal transformation \cite{Wald:1984rg},
\begin{align}
	\widehat R
	\,&\stackrel{\text{conf}}{=}\,
	\Omega ^{-2}[
		R - 2(n-1)g^{\mu \nu }\nabla _\mu \nabla _\nu \ln \sqrt{A}
        \\[0.5em]
        &\quad
		-\,
		(n-2)(n-1)g^{\mu \nu } (\nabla _\mu \ln \sqrt{A})(\nabla _\nu \ln \sqrt {A})
	].
    \nonumber
\end{align}
This implies that the Einstein-Hilbert action is in general, not invariant under conformal transformation. Nevertheless\footnote{On a related note, the fact that the Einstein-Hilbert action is topological in nature in $ n = 2 $ dimensions may lead one to introduce a (dilaton) field in a form coupled to $R$ and a corresponding (linear) potential as in the case of Jackiw-Teitelboim gravity\cite{Jackiw,Teitelboim}. This addresses the problem of the suppression of fluctuation in quantum gravity for $ n = 2 $ dimensions.}, for $n = 2$,
\begin{align}
	\widehat R
	\,&\stackrel{\text{conf}}{=}\,
	\frac{R}{A} 
	- 
	\frac{2}{A}\square (\ln A^{1/2}). 
\end{align}
Noting that $ \sqrt{-\widehat g} = A\sqrt{-g} $, the second term on the right hand side becomes simply a total derivative in the action. In effect, with the cancellation of the inverse conformal factor coefficient of the Ricci scalar upon multiplication by $\sqrt{-\widehat g}$, we find $\widehat S^{(2)} = S^{(2)}$, as concluded above.

\section{Summary and concluding remarks}
\label{seConclude}

\noindent
Our previous study has shown that the massless Klein-Gordon equation is invariant under the general disformal transformation subject to some conditions that we organised in the current work as the `triple KG constraints'. Given that the Klein-Gordon and Einstein-Hilbert actions are both special cases of the Horndeski action---the most general second-order scalar-tensor action that is form-invariant under the \emph{special disformal transformation}---it is interesting to consider the \textit{general} disformal transformation of the Einstein-Hilbert action under the triple KG constraints. Although the massless KG equation and the EH action are not exactly on equal footing within the context of Horndeski theory, our insight is that the triple KG constraints would significantly simplify the resulting transformation of the Einstein-Hilbert action. That is, with significant fragments from the Horndeski action (in line with another study on its \textit{special} disformal transformation) and manageably small number of `extraneous' or beyond-Horndeski terms.

As it turns out, the general disformal transformation of the $n$-dimensional Einstein-Hilbert action, subject to the triple KG constraints, leads to a sum of the original action and a disformal contribution involving only a six-term Lagrangian that includes the Ricci tensor coupled to a sum of derivatives of scalar fields and kinetic terms. Although the transformation does not lead to invariance in general spacetime dimensions, the number of `extraneous' terms at the level of $\Phi_\mu$ is significantly reduced when compared to that for unconstrained general disformal transformation. Furthermore, we identify two of the four Horndeski sub-Lagrangians resulting from the transformation. In four spacetime dimensions, we find the number of disformal contribution terms in the Lagrangian to be the same as that for $n > 4$. This decreases to a five-term Lagrangian from the original six-term disformal contributions, in three dimensions. Interestingly, in two spacetime dimensions, the Einstein-Hilbert action is invariant under the constrained disformal transformation!

Since the general disformal transformation subject to the triple KG constraints is invertible, the transformed action gains no additional degrees of freedom within the framework of the full Horndeski theory; that is, consistent with previous studies on primordial cosmological perturbations and investigations on theories beyond Horndeski. Nonetheless, for the Einstein-Hilbert action alone, the transformation leaves disformal contribution terms, in general spacetime dimensions, that may not be easily accounted for. Their complexity and the question of whether they can be related to pressing problems such as dark matter and dark energy, call for a separate study. In addition to this, the seemingly sudden `transition' from non-invariance to invariance under the constrained disformal transformation from high dimensions to $n = 2$, while considerably an interesting mathematical find on its own, deserves a deeper connection to the physical aspect of such `transition'. We hope to traverse these new avenues opened in the current work in our future studies.

\newpage
\section*{Appendix A: Unconstrained Disformal Transformation of the Ricci Tensor}
\label{AppendixA}
The unconstrained general disformal transformation of the Ricci tensor is given by
\begin{align}
    \label{hatRmnNocons}
    &\widehat{R}_{\mu\nu}\nonumber = R_{\mu\nu} 
    + F_1g_{\mu\nu}
    + \frac{[A(n-2)-2\chi(n-3)]}{2A(A-2\chi)}A_{;\mu\nu}\nonumber
    \\[0.5em]
    &+\, 
    \frac{[3A^2(n-2) + 12\chi^2 (n-3) - 4A\chi(3n-8)]}{4A^2(A-2\chi)^2}A_{;\mu}A_{;\nu}\nonumber
    \\[0.5em]
    &+\,
    F_2\Phi_{\mu}\Phi_{\nu} 
    + 
    F_3\Phi_{(\mu;\nu)} 
    + 
    F_4[\Phi_{(\mu}\chi_{;\nu)} 
    - 
    \Phi_{\mu}A_{;\nu}]\nonumber
    \\[0.5em]
    &+\,
    M^\alpha\Phi_{\alpha;(\mu}\Phi_{\nu)} 
    + 
    N^\alpha\Phi_{(\mu}\Phi_{\nu);\alpha} 
    + 
    \frac{\Phi^\alpha\chi_{;(\mu}\Phi_{\nu);\alpha}}{A(A-2\chi)}\nonumber
    \\[0.5em]
    &-\,
    \frac{\Phi^\alpha A_{;(\mu}\Phi_{\nu);\alpha}}{A(A-2\chi)} 
    + \frac{\Phi^\alpha\Phi^\beta\Phi_{\mu;\beta}\Phi_{\nu;\alpha}}{2A(A-2\chi)} 
    - 
    \frac{(A-\chi)\Phi_{\mu}{}^{;\alpha}\Phi_{\nu;\alpha}}{A(A-2\chi)}\nonumber
    \\[0.5em]
    &-\, \frac{\Phi^\alpha\Phi_{(\mu}A_{;\nu)\alpha}}{A(A-2\chi)} 
    + 
    \frac{\Phi_{\alpha;(\mu}\Phi_{\nu)}{}^{;\alpha}}{A} 
    - 
    \frac{2(A-\chi)A_{;(\mu}\chi_{;\nu)}}{A(A-2\chi)^2}\nonumber
    \\[0.5em]
    &+\, 
    \frac{(3A-2\chi)\chi_{;\mu}\chi_{;\nu}}{2A(A-2\chi)^2} 
    - 
    \frac{\Phi^\alpha{}_{;\mu}\Phi_{\alpha;\nu}}{A(A-2\chi)} 
    +\, 
    \frac{\Phi^\alpha\Phi_{(\mu;\nu)\alpha}}{A-2\chi}\nonumber
    \\[0.5em]
    &+\, \frac{\Phi^\alpha\Phi^\beta\Phi_{(\mu}\Phi_{\nu);\alpha\beta}}{A(A-2\chi)} 
    - 
    \frac{\Phi_{(\mu}\Phi_{\nu);\alpha}{}{}^\alpha}{A} 
    - \frac{\Phi^\alpha\Phi^\beta\Phi_{\alpha;\beta(\mu}\Phi_{\nu)}}{A(A-2\chi)}\nonumber
    \\[0.5em]
    &+\, \frac{\Phi^\alpha{}_{;\alpha(\mu}\Phi_{\nu)}}{A} - \frac{\Phi^\alpha\Phi_{\alpha;\mu\nu}}{A-2\chi},
\end{align}
where 
\begin{align}
    F_1 
    &= 
    -\frac{\Box A}{2A} 
    + 
    \frac{1}{2A(A-2\chi)}
    \bigg\{    
    A^\alpha\Phi^\beta\Phi_{\alpha;\beta} 
    + 
    \Phi^\alpha_{;\alpha}
    \Phi^\beta A_{;\beta}
    \nonumber
    \\[0.5em]
    &\qquad
    +\,
    \Phi^\alpha \Phi^\beta A_{;\alpha\beta}
    +
    A^{;\alpha}\chi_{;\alpha}  
    + \frac{\Phi^\alpha\chi_{;\alpha}\Phi^\beta A_{;\beta}}{A-2\chi}    
    \nonumber
    \\[0.5em]
    &\qquad
    - 
    \frac{[A(n-4) - 2\chi(n-5)]A^{;\alpha}A_{;\alpha}}{2A}
    \\[0.5em]
    &\qquad 
    +\, 
    \frac{[A(n-6) - 2\chi(n-5)]\Phi^\alpha A_{;\alpha}\Phi^\beta A_{;\beta}}{2A(A-2\chi)}\bigg\},
    \nonumber
    \\[0.5em]
    F_2 
    &= \frac{\Phi^{\alpha;\beta}\Phi_{[\alpha;\beta]}}{A^2} 
    + 
    \frac{1}{2A^2(A-2\chi)}\big(
    -\Phi^\alpha{}_{;\rho}\Phi_{\alpha;\beta}\Phi^{\beta}\Phi^\rho
    \nonumber
    \\[0.5em]
    &\qquad 
    -\, 
    2\chi^{;\alpha}\Phi_{;\alpha\beta}\Phi^\beta 
    + 
    2A^{;\alpha}\Phi^{\beta}\Phi_{\alpha;\beta} 
    - 
    \chi^{;\alpha}\chi_{;\alpha} 
    \nonumber
    \\[0.5em]
    &\qquad
    +\,
    2A^{;\alpha}X_{;\alpha} 
    - 
    A^{;\alpha}A_{;\alpha}\big),
    \\[0.5em]
    F_3 
    &= 
    \frac{1}{A-2\chi}
    \bigg\{
    \frac{[(n-4)A - 2(n-3)\chi]}{2A(A-2\chi)}\Phi^{;\alpha}A_{;\alpha}
    \nonumber
    \\[0.5em]
    &\qquad\qquad\qquad
    +\,
    {\Phi^\alpha}_{;\alpha}
    + 
    \frac{\Phi^\alpha\chi_{;\alpha}}{A-2\chi}
    \bigg\},
    \\[0.5em]
    F_4 
    &= 
    \frac{1}{A(A-2\chi)}
    \bigg\{ 
    \frac{[(n-6)A - 2(n-5)\chi]}
    {2A(A-2\chi)}
    \Phi^\alpha A_{;\alpha}
    \nonumber
    \\[0.5em]
    &\qquad\qquad\qquad 
    +\,
    {\Phi^\alpha}_{;\alpha}
    +
    \frac{\Phi^\alpha\chi_{;\alpha}}{A-2\chi} 
    \bigg\},
    \\[0.5em]
    M^\alpha 
    &= 
    \frac{1}{A(A-2\chi)}
    \bigg\{
    \frac{[(n-4)A - 2(n-5)\chi]}{2A}
    A^{;\alpha}
    \nonumber
    \\[0.5em]
    &\qquad\qquad\qquad 
    -\Phi^{\alpha;\beta}\Phi_{\beta} 
    - 
    \chi^{;\alpha}
    \bigg\},
    \\[0.5em]
    N^\alpha 
    &= 
    \frac{1}{A(A-2\chi)}
    \bigg\{ 
    \frac{\Phi^\alpha\Phi^\beta \chi_{;\beta}}
    {A-2\chi} 
    + 
    \Phi^\alpha\Phi^\beta{}_{;\beta}
    \\[0.5em]
    & 
    +\, 
    [(n-6)A-2(n-5)\chi]
    \bigg[
    \frac{\Phi^\alpha\Phi^\beta A_{;\beta}}
    {{2A(A-2\chi)}}
    - 
    \frac{A^{;\alpha}}{2A}
    \bigg]\bigg\}.
    \nonumber
\end{align}
These ``new'' terms stem out from parts of \eqref{KGFS3} that were annihilated by the Klein-Gordon constraints.

\section*{Appendix B: Unconstrained Disformal Transformation of the Ricci Scalar and the Action}
\label{AppendixB}

\noindent
The unconstrained general disformal transformation of the Ricci scalar is given by 
\begin{align}
    \widehat{R}&= \frac{R}{A} - \frac{R_{\mu\nu}\Phi^\mu\Phi^\nu}{A(A-2\chi)} + \tilde{\mathcal{U}},\label{RGDT}
\end{align}
where
\begin{align}
    \tilde{\mathcal{U}} &= \frac{(\Phi^\alpha{}_{;\alpha})^2}{A(A-2\chi)} - \frac{(2A-\chi)\Phi^{\alpha;\beta}\Phi_{\alpha;\beta}}{A^2(A-2\chi)}\nonumber\\
    &\qquad +\, \frac{(A-\chi)\Phi^{\alpha;\beta}\Phi_{\beta;\alpha}}{A^2(A-2\chi)} + \frac{2\Phi^\alpha{}_{;\alpha}\Phi^\beta\chi_{;\beta}}{A(A-2\chi)^2}\nonumber\\
    &\qquad +\, \frac{[A(n-3) - 2\chi(n-2)]\Phi^{\alpha}{}_{;\alpha}\Phi^\beta A_{;\alpha}}{A^2(A-2\chi)^2}\nonumber\\
    &\qquad +\, \frac{(n-2)\Phi^\alpha\chi_{;\alpha}\Phi^\beta A_{;\beta}}{A^2(A-2\chi)^2} + \frac{(5A-2\chi)\chi^{;\alpha}\chi_{;\alpha}}{2A^2(A-2\chi)^2}\nonumber\\
    &\qquad -\, \frac{[A(n-9) - 2\chi(n-7)](n-2)(\Phi^\alpha A_{;\alpha})^2}{4A^3(A-2\chi)^2}\nonumber\\
    &\qquad +\, \frac{[A(n-5) - 2\chi(n-3)]A^{;\alpha}\chi_{;\alpha}}{A^2(A-2\chi)^2}\nonumber\\
    &\qquad -\, \frac{A^{;\alpha}A_{;\alpha}}{4A^3(A-2\chi)^2}\bigg[A^2(n-6)(n-1)\nonumber\\
    &\qquad\qquad\qquad\qquad\qquad  +\, 4\chi^2(n-7)(n-2)\nonumber\\
    &\qquad\qquad\qquad\qquad\qquad  -\, 4A\chi(n^2-8n + 11)\bigg]\nonumber\\
    &\qquad - \,\frac{[A(n-1) - 2\chi(n-2)]}{A^2(A-2\chi)}\Box A\nonumber\\
    &\qquad +\, \frac{\Phi_{\alpha;\beta}\chi^{;\alpha}\Phi^\beta}{A^2(A-2\chi)} + \frac{(n-3)\Phi_{\alpha;\beta}A^{;\alpha}\Phi^\beta}{A^2(A-2\chi)}\nonumber\\
    &\qquad +\, \frac{(n-2)A_{;\alpha\beta}\Phi^\alpha\Phi^\beta}{A^2(A-2\chi)} + \frac{\Phi^{\alpha;\mu}\Phi_{\alpha;\beta}\Phi^\beta\Phi_\mu}{2A^2(A-2\chi)}\nonumber\\
    &\qquad +\, \frac{2(\Phi^\alpha{}_{;\alpha\beta}\Phi^\beta - \Phi_{\beta;\alpha}{}{}^\alpha\Phi^\beta)}{A(A-2\chi)} \label{RExtraGDT}
\end{align}

It follows from \eqref{RGDT}, \eqref{RExtraGDT}, and \eqref{gdetGDT} that the unconstrained general disformal transformation of the Einstein-Hilbert action is 
\begin{align}
    \widehat{S} 
    &= 
    \frac{1}{2}\int d^nx\sqrt{-g} \sqrt{A^{n-1}(A-2\chi)}\nonumber\\
    &\qquad\qquad\qquad \times\, \bigg[\frac{R}{A} - \frac{R_{\mu\nu}\Phi^\mu\Phi^\nu}{A(A-2\chi)} + \mathcal{U}\bigg]\label{UnconstEHA}
\end{align}
There are two second-order derivatives of $\Phi_\mu$ in \eqref{UnconstEHA}. One is $2A^{(n-3)/2}\Phi^\alpha{}_{;\alpha\beta}\Phi^\beta/\sqrt{A-2\chi}$, while the other is 
$-2A^{(n-3)/2}\Phi_{\beta;\alpha}{}^{\alpha}\Phi^\beta/\sqrt{A-2\chi}$
With the transformed Ricci scalar inside the action integral, we may now integrate by parts these higher-order terms:
\begin{align}
    &\frac{2A^{(n-3)/2}\Phi^\alpha{}_{;\alpha\beta}\Phi^\beta}{\sqrt{A-2\chi}} =\nonumber\\
    & -\,\frac{2A^{(n-3)/2}\Phi^\alpha{}_{;\alpha}\Phi^\beta\chi_{;\beta}}{(A-2\chi)^{3/2}} - \frac{2A^{(n-3)/2}\Phi^\alpha{}_{;\alpha}\Phi^\beta{}_{;\beta}}{\sqrt{A-2\chi}}\nonumber\\
    &-\, \frac{A^{(n-5)/2}[A(n-4) - 2\chi(n-3)]}{(A-2\chi)^{3/2}}\Phi^\alpha{}_{;\alpha}\Phi^\beta A_{;\beta}\\[0.5em]
    &-\frac{2A^{(n-3)/2}\Phi_{\beta;\alpha}{}^{\alpha}\Phi^\beta}{\sqrt{A-2\chi}}=\nonumber\\
    & +\, \frac{2A^{(n-3)/2}\Phi^\alpha\chi^{;\beta}\Phi_{\alpha;\beta}}{(A-2\chi)^{3/2}} + \frac{2A^{(n-3)/2}\Phi^{\alpha;\beta}\Phi_{\alpha;\beta}}{\sqrt{A-2\chi}}\nonumber\\
    &+\, \frac{A^{(n-5)/2}[A(n-4)-2\chi(n-3)]}{(A-2\chi)^{3/2}}\Phi^\alpha A^{;\beta}\Phi_{\alpha;\beta}
\end{align}
These generated terms combine with those already in $\mathcal{U}$. The transformed action \eqref{UnconstEHA} then becomes
\begin{align}
    \label{unconstrhatS}
    \widehat{S} 
    &= 
    \frac{1}{2}\int \dd^nx\sqrt{-g} \bigg(A^{(n-3)/2}
    \sqrt{A-2\chi}R
    \nonumber
    \\[0.5em]
    &\qquad\qquad-\, \frac{A^{(n-3)/2}\Phi^\mu\Phi^\nu R_{\mu\nu}}{\sqrt{A-2\chi}} + \tilde{\mathcal{H}}\bigg),
\end{align}
where
\begin{align}
    \label{tildeH}
    \tilde{\mathcal{H}} 
    &= 
    -\frac{A^{(n-3)/2}(\Phi^\alpha{}_{;\alpha})^2}{\sqrt{A-2\chi}} + \frac{A^{(n-5)/2}\chi}{\sqrt{A-2\chi}}\Phi^{\alpha;\beta}\Phi_{\alpha;\beta}
    \nonumber
    \\
    &
    +\, 
    \frac{A^{(n-5)/2}(A-\chi)}{\sqrt{A-2\chi}}
    \Phi^{\alpha;\beta}\Phi_{\beta;\alpha}
    \nonumber
    \\[0.5em]
    &+\, 
    \frac{A^{(n-5)/2}\Phi^{\alpha}{}_{;\alpha}\Phi^{\beta}A_{;\beta}}
    {\sqrt{A-2\chi}} 
    + 
    \frac{A^{(n-5)/2}\Phi^\alpha\chi_{;\alpha}\Phi^\beta A_{;\beta}}
    {(A-2\chi)^{3/2}}
    \nonumber
    \\[0.5em]
    &+\, 
    \frac{A^{(n-5)/2}[A(n-9) - 2\chi(n-7)](n-2)}
    {4(A-2\chi)^{3/2}}(\Phi^\alpha A_{;\alpha})^2
    \nonumber
    \\[0.5em]
    & 
    +\, 
    \frac{A^{(n-5)/2}(5A-2\chi)}{2(A-2\chi)^{3/2}}
    \chi^{;\alpha}\chi_{;\alpha}
    \nonumber\\[0.5em]
    &+\, 
    \frac{A^{(n-5)/2}[A(n-5) - 2\chi(n-3)]}
    {(A-2\chi)^{3/2}}A^{;\alpha}\chi_{;\alpha}
    \nonumber
    \\[0.5em]
    &-\, 
    \frac{A^{(n-7)/2}A^{;\alpha}A_{;\alpha}}{4(A-2\chi)^{3/2}}
    \big[A^2(n-6)(n-1)
    \nonumber
    \\[0.5em]
    &\qquad
    +\, 4\chi^2(n-7)(n-2)
    -\, 4A\chi(n^2-8n + 11)\big]
    \nonumber
    \\[0.5em]
    &-\, 
    \frac{A^{(n-5)/2}[A(n-1) - 2\chi(n-2)]}{\sqrt{A-2\chi}}\Box A
    \nonumber
    \\[0.5em]
    &+\, 
    \frac{2A^{(n-3)/2}\Phi^\alpha
    \chi^{;\beta}\Phi_{\alpha;\beta}}
    {(A-2\chi)^{3/2}} 
    + 
    \frac{A^{(n-5)/2}\chi^{;\alpha}
    \Phi^\beta\Phi_{\alpha;\beta}}{\sqrt{A-2\chi}}
    \nonumber
    \\[0.5em]
    &+\, 
    \frac{A^{(n-5)/2}(n-2)}{\sqrt{A-2\chi}}
    \Phi^\alpha\Phi^\beta A_{;\alpha\beta}
    \nonumber
    \\[0.5em]
    &+\, 
    \frac{A^{(n-5)/2}(n-3)}
    {\sqrt{A-2\chi}}A^{;\alpha}\Phi^\beta\Phi_{\alpha;\beta}
    \nonumber
    \\[0.5em]
    &+\, 
    \frac{A^{(n-5)/2}[A(n-4) - 2\chi(n-3)]}
    {(A-2\chi)^{3/2}}\Phi^\alpha A^{;\beta}\Phi_{\alpha,\beta}
    \nonumber
    \\[0.5em]
    &+\, 
    \frac{A^{(n-5)/2}\Phi^{\alpha;\mu}
    \Phi_{\alpha;\beta}\Phi^\beta\Phi_\mu}{2\sqrt{A-2\chi}}
\end{align}
\end{multicols}

\begin{thebibliography}{0}    

    \bibitem{Bekenstein:1992pj}
    J.~D.~Bekenstein,
    ``The Relation between physical and gravitational geometry,''
    Phys. Rev. D \textbf{48} (1993), 3641-3647
    doi:10.1103/PhysRevD.48.3641
    [arXiv:gr-qc/9211017 [gr-qc]].
    
    \bibitem{Bettoni:2013diz}
    D.~Bettoni and S.~Liberati,
    ``Disformal invariance of second order scalar-tensor theories: Framing the Horndeski action,''
    Phys. Rev. D \textbf{88} (2013), 084020
    doi:10.1103/PhysRevD.88.084020
    [arXiv:1306.6724 [gr-qc]].
    
    \bibitem{Kobayashi:2019hrl}
    T.~Kobayashi, ``Horndeski theory and beyond: a review,'' Rept. Prog. Phys. \textbf{82} (2019) no.8, 086901 doi:10.1088/1361-6633/ab2429
    [arXiv:1901.07183 [gr-qc]]. 

    \bibitem{Tsujikawa:2014uza}
    S.~Tsujikawa,
    ``Disformal invariance of cosmological perturbations in a generalized class of Horndeski theories,''
    JCAP \textbf{04} (2015), 043
    doi:10.1088/1475-7516/2015/04/043
    [arXiv:1412.6210 [hep-th]].
    
    \bibitem{Domenech:2015hka}
    G.~Dom\`enech, A.~Naruko and M.~Sasaki,
    ``Cosmological disformal invariance,''
    JCAP \textbf{10} (2015), 067
    doi:10.1088/1475-7516/2015/10/067
    [arXiv:1505.00174 [gr-qc]].
    
    \bibitem{Motohashi:2015pra}
    H.~Motohashi and J.~White,
    ``Disformal invariance of curvature perturbation,''
    JCAP \textbf{02} (2016), 065
    doi:10.1088/1475-7516/2016/02/065
    [arXiv:1504.00846 [gr-qc]].

    \bibitem{Alinea:2020laa}
    A.~L.~Alinea and T.~Kubota,
    ``Transformation of primordial cosmological perturbations under the general extended disformal transformation,''
    Int. J. Mod. Phys. D \textbf{30} (2021) no.08, 2150057
    doi:10.1142/S0218271821500577
    [arXiv:2005.12747 [gr-qc]].

    \bibitem{Makino:1991sg}
    N.~Makino and M.~Sasaki,
    ``The Density perturbation in the chaotic inflation with nonminimal coupling,''
    Prog. Theor. Phys. \textbf{86} (1991), 103-118
    doi:10.1143/PTP.86.103

    \bibitem{Fakir:1990eg}
    R.~Fakir and W.~G.~Unruh,
    ``Improvement on cosmological chaotic inflation through nonminimal coupling,''
    Phys. Rev. D \textbf{41} (1990), 1783-1791
    doi:10.1103/PhysRevD.41.1783

    \bibitem{Alinea:2015pza}
    A.~L.~Alinea, T.~Kubota, Y.~Nakanishi and W.~Naylor,
    ``Adiabatic regularisation of power spectra in $k$-inflation,''
    JCAP \textbf{06} (2015), 019
    doi:10.1088/1475-7516/2015/06/019
    [arXiv:1503.08073 [gr-qc]].
    
    \bibitem{Alinea:2016qlf}
    A.~L.~Alinea,
    ``Adiabatic regularization of power spectra in nonminimally coupled chaotic inflation,''
    JCAP \textbf{10} (2016), 027
    doi:10.1088/1475-7516/2016/10/027
    [arXiv:1607.05328 [gr-qc]].
    
    \bibitem{Alinea:2017ncx}
    A.~L.~Alinea and T.~Kubota,
    ``Adiabatic regularization of the power spectrum in nonminimally coupled general single-field inflation,''
    Phys. Rev. D \textbf{97} (2018) no.6, 063513
    doi:10.1103/PhysRevD.97.063513
    [arXiv:1709.06450 [gr-qc]].

    \bibitem{Sato:2017qau}
    S.~Sato and K.~i.~Maeda,
    ``Hybrid Higgs Inflation: The Use of Disformal Transformation,''
    Phys. Rev. D \textbf{97} (2018) no.8, 083512 doi:10.1103/PhysRevD.97.083512
    [arXiv:1712.04237 [gr-qc]].   

    \bibitem{Kaloper:2003yf}
    N.~Kaloper, ``Disformal inflation,''
    Phys. Lett. B \textbf{583} (2004), 1-13
    doi:10.1016/j.physletb.2004.01.005
    [arXiv:hep-ph/0312002 [hep-ph]]. 
    
    
	\bibitem{Gialamas:2024jeb}
	I.~D.~Gialamas, T.~Katsoulas and K.~Tamvakis,
	``Inflation and reheating in quadratic metric-affine gravity with derivative couplings,''
	JCAP \textbf{06} (2024), 005
	doi:10.1088/1475-7516/2024/06/005
	[arXiv:2403.08530 [gr-qc]].   
	
	\bibitem{Gialamas:2020vto}
	I.~D.~Gialamas, A.~Karam, A.~Lykkas and T.~D.~Pappas,
	``Palatini-Higgs inflation with nonminimal derivative coupling,''
	Phys. Rev. D \textbf{102} (2020) no.6, 063522
	doi:10.1103/PhysRevD.102.063522
	[arXiv:2008.06371 [gr-qc]]. 
	
	\bibitem{Gialamas:2021enw}
	I.~D.~Gialamas, A.~Karam, T.~D.~Pappas and V.~C.~Spanos,
	``Scale-invariant quadratic gravity and inflation in the Palatini formalism,''
	Phys. Rev. D \textbf{104} (2021) no.2, 023521
	doi:10.1103/PhysRevD.104.023521
	[arXiv:2104.04550 [astro-ph.CO]].
    
    \bibitem{Falciano:2011rf}
    F.~T.~Falciano and E.~Goulart,
    ``A new symmetry of the relativistic wave equation,'' Class. Quant. Grav. \textbf{29} (2012), 085011 doi:10.1088/0264-9381/29/8/085011
    [arXiv:1112.1341 [gr-qc]].
    
    \bibitem{Alinea:2022ygr}
    A.~L.~Alinea and M.~R.~D.~Chua,
    ``Extending the symmetry of the massless Klein\textendash{}Gordon equation under the general disformal transformation,''
    Int. J. Mod. Phys. A \textbf{38} (2023) no.01, 2350004
    doi:10.1142/S0217751X23500045
    [arXiv:2208.10953 [gr-qc]].

    \bibitem{Goulart:2013laa}
    E.~Goulart and F.~T.~Falciano,
    ``Disformal invariance of Maxwell's field equations,''
    Class. Quant. Grav. \textbf{30} (2013), 155020
    doi:10.1088/0264-9381/30/15/155020
    [arXiv:1303.4350 [gr-qc]].
    
    \bibitem{Bittencourt:2015ypa}
    E.~Bittencourt, I.~P.~Lobo and G.~G.~Carvalho,
    ``On the disformal invariance of the Dirac equation,''
    Class. Quant. Grav. \textbf{32} (2015), 185016
    doi:10.1088/0264-9381/32/18/185016
    [arXiv:1505.03415 [gr-qc]].

    \bibitem{Alinea:2020sei}
    A.~L.~Alinea,
    ``On the Disformal Transformation of the Einstein-Hilbert Action,'' Jordan J. Phys. \textbf{16} (2024) no. 5 506-515
    doi:10.47011/16.5
    [arXiv:2010.00956 [gr-qc]].  

    \bibitem{Gleyzes:2014qga}
    J.~Gleyzes, D.~Langlois, F.~Piazza and F.~Vernizzi, ``Exploring gravitational theories beyond Horndeski,''
    JCAP \textbf{02} (2015), 018
    doi:10.1088/1475-7516/2015/02/018
    [arXiv:1408.1952 [astro-ph.CO]].

    \bibitem{Langlois:2015cwa}
    D.~Langlois and K.~Noui,
    ``Degenerate higher derivative theories beyond Horndeski: evading the Ostrogradski instability,''
    JCAP \textbf{02} (2016), 034
    doi:10.1088/1475-7516/2016/02/034
    [arXiv:1510.06930 [gr-qc]].  


    \bibitem{Chatzifotis:2021hpg} 
    N.~Chatzifotis, E.~Papantonopoulos and C.~Vlachos,
    ``Disformal transition of a black hole to a wormhole in scalar-tensor Horndeski theory,''
    Phys. Rev. D \textbf{105} (2022) no.6, 064025
    doi:10.1103/PhysRevD.105.064025
    [arXiv:2111.08773 [gr-qc]]. 

    \bibitem{BenAchour:2020wiw}
    J.~Ben Achour, H.~Liu and S.~Mukohyama,
    ``Hairy black holes in DHOST theories: Exploring disformal transformation as a solution-generating method,''
    JCAP \textbf{02} (2020), 023
    doi:10.1088/1475-7516/2020/02/023
    [arXiv:1910.11017 [gr-qc]]. 

    \bibitem{Domenech:2023ryc}
    G.~Dom\`enech and A.~Ganz,
    ``Disformal symmetry in the Universe: mimetic gravity and beyond,''
    JCAP \textbf{08} (2023), 046
    doi:10.1088/1475-7516/2023/08/046
    [arXiv:2304.11035 [gr-qc]].  

	\bibitem{Hell:2024xbv}
	A.~Hell,
	``Unveiling the inconsistency of the Proca theory with non-minimal coupling to gravity,''
	[arXiv:2403.18673 [gr-qc]].
    
    \bibitem{Takahashi:2021ttd}
    K.~Takahashi, H.~Motohashi and M.~Minamitsuji, ``Invertible disformal transformations with higher derivatives,''
    Phys. Rev. D \textbf{105} (2022) no.2, 024015 doi:10.1103/PhysRevD.105.024015
    [arXiv:2111.11634 [gr-qc]]. 

    \bibitem{Takahashi:2022mew}
    K.~Takahashi, M.~Minamitsuji and H.~Motohashi, ``Generalized disformal Horndeski theories: Cosmological perturbations and consistent matter coupling'', PTEP \textbf{2023} (2023) no.1, 013E01
    doi:10.1093/ptep/ptac161
    [arXiv:2209.02176 [gr-qc]].
    
    \bibitem{Woodard:2015zca}
    R.~P.~Woodard, ``Ostrogradsky's theorem on Hamiltonian instability,''
    Scholarpedia \textbf{10} (2015) no.8, 32243 doi:10.4249/scholarpedia.32243
    [arXiv:1506.02210 [hep-th]].

    \bibitem{Goulart:2020wkq}
    \'E.~Goulart and E.~Bittencourt,
    ``Space and time ambiguities in vacuum electrodynamics,''
    Class. Quant. Grav. \textbf{38}, no.14, 145029 (2021)
    doi:10.1088/1361-6382/ac08a9
    [arXiv:2010.14936 [gr-qc]].

    \bibitem{Gleyzes:2014dya}
    J.~Gleyzes, D.~Langlois, F.~Piazza and F.~Vernizzi,
    ``Healthy theories beyond Horndeski,''
    Phys. Rev. Lett. \textbf{114} (2015) no.21, 211101
    doi:10.1103/PhysRevLett.114.211101
    [arXiv:1404.6495 [hep-th]].

    \bibitem{BenAchour:2016cay}
    J.~Ben Achour, D.~Langlois and K.~Noui,
    ``Degenerate higher order scalar-tensor theories beyond Horndeski and disformal transformations,''
    Phys. Rev. D \textbf{93} (2016) no.12, 124005
    doi:10.1103/PhysRevD.93.124005
    [arXiv:1602.08398 [gr-qc]].

    \bibitem{Galtsov:2018xuc}
    D.~Gal'tsov and S.~Zhidkova,
    ``Ghost-free Palatini derivative scalar\textendash{}tensor theory: Desingularization and the speed test,''
    Phys. Lett. B \textbf{790} (2019), 453-457
    doi:10.1016/j.physletb.2019.01.061
    [arXiv:1808.00492 [hep-th]].
    
	\bibitem{Alinea:2024jrf}
	A.~L.~Alinea and J.~D.~Ordonez,
	``On the KG-constrained Bekenstein's disformal transformation of the Einstein-Hilbert action,''
	[arXiv:2408.06915 [gr-qc]].    

    \bibitem{Minamitsuji:2014waa}
    M.~Minamitsuji,
    ``Disformal transformation of cosmological perturbations,''
    Phys. Lett. B \textbf{737}, 139-150 (2014)
    doi:10.1016/j.physletb.2014.08.037
    [arXiv:1409.1566 [astro-ph.CO]].
    
    \bibitem{Horndeski:1974wa}
    G.~W.~Horndeski, ``Second-order scalar-tensor field equations in a four-dimensional space,''
    Int. J. Theor. Phys. \textbf{10} (1974), 363-384 doi:10.1007/BF01807638    

    \bibitem{Wald:1984rg}
    R.~M.~Wald,
    ``General Relativity,''
    Chicago Univ. Pr., 1984,
    doi:10.7208/chicago/9780226870373.001.0001


    \bibitem{Fujita:2015ymn}
    T.~Fujita, X.~Gao and J.~Yokoyama,
    ``Spatially covariant theories of gravity: disformal transformation, cosmological perturbations and the Einstein frame,''
    JCAP \textbf{02}, 014 (2016)
    doi:10.1088/1475-7516/2016/02/014
    [arXiv:1511.04324 [gr-qc]].


    \bibitem{Carvalho:2017obq}
    G.~G.~Carvalho, E.~Bittencourt and I.~P.~Lobo,
    ``On the disformal invariance of the massless Dirac equation,''
     \textit{Proceedings, 14th Marcel Grossmann Meeting on Recent Developments in Theoretical and Experimental General Relativity, Astrophysics, and Relativistic Field Theories} 2648-2653 (2017),
    doi:10.1142/9789813226609\_0322

    \bibitem{Sakstein:2014isa}
    J.~Sakstein,
    ``Disformal Theories of Gravity: From the Solar System to Cosmology,''
    JCAP \textbf{12}, 012 (2014)
    doi:10.1088/1475-7516/2014/12/012
    [arXiv:1409.1734 [astro-ph.CO]].

    \bibitem{vandeBruck:2015ida}
    C.~van de Bruck and J.~Morrice,
    ``Disformal couplings and the dark sector of the universe,''
    JCAP \textbf{04}, 036 (2015)
    doi:10.1088/1475-7516/2015/04/036
    [arXiv:1501.03073 [gr-qc]].

    \bibitem{Jirousek:2022rym}
    P.~Jirou\v{s}ek, K.~Shimada, A.~Vikman and M.~Yamaguchi,
    ``Disforming to conformal symmetry,''
    JCAP \textbf{11}, 019 (2022)
    doi:10.1088/1475-7516/2022/11/019
    [arXiv:2207.12611 [gr-qc]].

    \bibitem{Zumalacarregui:2010wj}
    M.~Zumalacarregui, T.~S.~Koivisto, D.~F.~Mota and P.~Ruiz-Lapuente,
    ``Disformal Scalar Fields and the Dark Sector of the Universe,''
    JCAP \textbf{05}, 038 (2010)
    doi:10.1088/1475-7516/2010/05/038
    [arXiv:1004.2684 [astro-ph.CO]].

    \bibitem{Zumalacarregui:2013pma}
    M.~Zumalac\'arregui and J.~Garc\'\i{}a-Bellido,
    ``Transforming gravity: from derivative couplings to matter to second-order scalar-tensor theories beyond the Horndeski Lagrangian,''
    Phys. Rev. D \textbf{89}, 064046 (2014)
    doi:10.1103/PhysRevD.89.064046
    [arXiv:1308.4685 [gr-qc]].

    \bibitem{Crisostomi:2016czh}
    M.~Crisostomi, K.~Koyama and G.~Tasinato,
    ``Extended Scalar-Tensor Theories of Gravity,''
    JCAP \textbf{04}, 044 (2016)
    doi:10.1088/1475-7516/2016/04/044
    [arXiv:1602.03119 [hep-th]].

    \bibitem{Ikeda:2023ntu}
    T.~Ikeda, K.~Takahashi and T.~Kobayashi,
    ``Consistency of higher-derivative couplings to matter fields in scalar-tensor gravity,''
    Phys. Rev. D \textbf{108}, no.4, 044006 (2023) doi:10.1103/PhysRevD.108.044006
    [arXiv:2302.03418 [gr-qc]].

    \bibitem{Takahashi:2022ctx}
    K.~Takahashi, R.~Kimura and H.~Motohashi,
    ``Consistency of matter coupling in modified gravity,''
    Phys. Rev. D \textbf{107}, no.4, 044018 (2023)
    doi:10.1103/PhysRevD.107.044018
    [arXiv:2212.13391 [gr-qc]].

    \bibitem{Hilbert:1915tx}
    D.~Hilbert,
    ``Die Grundlagen der Physik. 1.,''
    Gott. Nachr. \textbf{27}, 395-407 (1915)

    
    \bibitem{Ishak:2018his}
    M.~Ishak,
    ``Testing General Relativity in Cosmology,''
    Living Rev. Rel. \textbf{22} (2019) no.1, 1
    doi:10.1007/s41114-018-0017-4
    [arXiv:1806.10122 [astro-ph.CO]].

    \bibitem{Robertson:1933zz}
    H.~P.~Robertson,
    ``Relativistic Cosmology,''
    Rev. Mod. Phys. \textbf{5} (1933), 62-90
    doi:10.1103/RevModPhys.5.62

	\bibitem{BransDicke}
	C.~Brans and R.~H.~Dicke,
	``Mach's principle and a relativistic theory of gravitation,''
	Phys. Rev. \textbf{124} (1961), 925-935
	doi:10.1103/PhysRev.124.925
	
	\bibitem{Jackiw}
	R.~Jackiw,
	``Lower Dimensional Gravity,''
	Nucl. Phys. B \textbf{252} (1985), 343-356
	doi:10.1016/0550-3213(85)90448-1
	
	\bibitem{Teitelboim}
	C.~Teitelboim,
	``Gravitation and Hamiltonian Structure in Two Space-Time Dimensions,''
	Phys. Lett. B \textbf{126} (1983), 41-45
	doi:10.1016/0370-2693(83)90012-6

\end{thebibliography}
\end{document}